\documentclass[acmtog, authorversion]{acmart}

\usepackage{booktabs} 
\usepackage{amsfonts}
\usepackage{amssymb}
\usepackage{amsmath}
\usepackage{wrapfig}
\usepackage{hyperref}
\usepackage{wrapfig}
\usepackage{nicefrac}

\setcitestyle{square}

\usepackage[ruled]{algorithm2e} 

\SetAlFnt{\small}
\SetAlCapFnt{\small}
\SetAlCapNameFnt{\small}
\SetAlCapHSkip{0pt}
\IncMargin{-\parindent}	

\acmJournal{TOG}

\newcommand{\nico}	[1]{{\color{orange}Nico: #1}}
\newcommand{\paolo}[1]{{\color{blue}Paolo: #1}}
\newcommand{\alla}	[1]{{\color{cyan}	Alla: #1}}

\definecolor{darkpastelgreen}{rgb}{0.01, 0.75, 0.24}
\newcommand{\enrico}	[1]{{\color{darkpastelgreen}Enrico: #1}}

\newcommand{\com} [1]{{}} 

\newcommand{\ignore} [1]{{}} 

\newcommand{\SM}{{\mathcal{M}}}				
\newcommand{\SMT}{{\mathcal{M}_2}}			
\newcommand{\SMF}{{\mathcal{M}_4}}			
\newcommand{\MM}{{\mathcal{MM}}}			
\newcommand{\VM}{{\mathcal{MV}}}				
\renewcommand{\L}{\mathcal{L}}				
\renewcommand{\G}{\mathcal{G}}				
\newcommand{\Pool}{\mathcal{P}}				
\newcommand{\PSeed}{\mathbf{P}}				
\newcommand{\CF}{\mathcal{CF}}				
\newcommand{\LL}{\mathbf{L}}					
\newcommand{\Li}[1]{\ell_{#1}}					
\newcommand{\Lo}{\ell}						
\newcommand{\Lf}{f}						
\newcommand{\X}{\mathbf{X}}					
\newcommand{\pz}{p_0}						
\newcommand{\ppt}{p_{\frac{\pi}{2}}}				
\newcommand{\pp}{p_{\pi}}
\newcommand{\ptp}{p_{\frac{3}{2}\pi}}
\newcommand{\ptheta}{p_{\theta}}

\begin{document}
\title{Loopy Cuts: Surface-Field Aware Block Decomposition for Hex-Meshing}

\author{Marco Livesu$^*$}
\affiliation{%
  \institution{CNR IMATI}
  \city{Genoa}
  \country{Italy}}
\email{marco.livesu@gmail.com}

\author{Nico Pietroni$^*$}
\affiliation{%
  \institution{University of Technology Sydney}
  \city{Sydney}
  \country{Australia}}

\author{Enrico Puppo}
\affiliation{%
  \institution{University of Genoa}
  \city{Genoa}
  \country{Italy}}
  
  \author{Alla Sheffer}
\affiliation{%
  \institution{University of British Columbia}
  \city{Vancouver}
  \country{Canada}}
  
  \author{Paolo Cignoni}
\affiliation{%
  \institution{CNR ISTI}
  \city{Pisa}
  \country{Italy}\\$^*$: joint first authors}

%


\begin{teaserfigure}
\includegraphics[width=\linewidth]{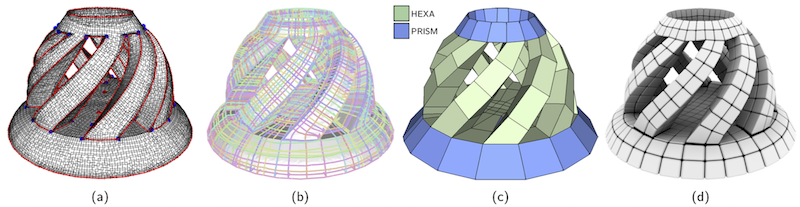}
\caption{We take as input a surface mesh together with a set of feature curves and a smooth surface field that aligns to them (a); we generate a pool of field-aware loops over it (b); and decompose the model into blocks by cutting it using surfaces bounded by algorithmically selected loops (c);  we produce a hexahedral mesh from this block decomposition via midpoint subdivision and mesh optimization (d).}
\label{fig:teaser}
\end{teaserfigure}

\begin{abstract}
We present a new fully automatic block-decomposition hexahedral meshing algorithm capable of producing high quality meshes that strictly preserve feature curve networks on the input surface and align with an input surface cross-field.  
We produce all-hex meshes on the vast majority of inputs, and introduce localized non-hex elements only when the surface feature network necessitates those.  
The input to our framework is a closed surface with a collection of geometric or user-demarcated feature curves and a feature-aligned surface cross-field.
Its output is a compact set of blocks whose edges interpolate these features and are loosely aligned with this cross-field. 
We obtain this block decomposition by cutting the input model using a collection of simple cutting surfaces bounded by closed surface loops. 
The set of cutting loops spans the input feature curves, ensuring feature preservation, and is obtained using a field-space sampling process. The computed loops are uniformly distributed across the surface, cross orthogonally, and are loosely aligned with the cross-field directions, inducing the desired block decomposition.
We validate our method by applying it to a large range of complex inputs and comparing our results to those produced by state-of-the-art alternatives. Contrary to prior approaches, our framework consistently produces high-quality field aligned meshes while strictly preserving geometric or user-specified surface features.
\end{abstract}

%
%

%
%


\maketitle


\section{Introduction}
\label{sec:intro}

Hexahedral and hex-dominant volumetric meshing of 3D shapes is a well investigated yet still open research topic. At their core, hexahedral meshing algorithms balance fidelity to the input surface geometry against element quality, and seek to compute meshes with well shaped, or as cuboid as possible, elements whose outer surface closely aligns with that of the input model (Section~\ref{sec:related}). 
To achieve high surface fidelity and to keep the element budget low, users prefer meshes whose external quads align with the surface curvature directions. Users also often strongly prefer meshes whose edges interpolate geometric or semantic surface feature curves.  We propose {\em Loopy Cuts}, a robust new meshing algorithm that is specifically designed to satisfy these preferences. 

Current attempts at automatic hex mesh generation fail to provide the combination of robustness, feature and curvature alignment we seek (Section~\ref{sec:related}). 
Methods based on adaptive grids~\cite{Marechal2009} or PolyCube maps~\cite{Livesu:2013:PolyCut} do not effectively handle surface features that do not align with the global frame, and cannot easily support alignment to a curvature field (Figure~\ref{fig:polycube_comparison}). Methods that align mesh elements with a volumetric frame field generate meshes that support feature and field alignment~\cite{liu2018singularity,Li:2012} but lack robustness, failing to automatically mesh more complex inputs (Section~\ref{sec:results}, Figure~\ref{fig:hand_joint}).


Our method complements these existing approaches by providing the desired combination of alignment and robustness. It achieves this goal by employing a meshing strategy that mimics 
semi-manual block decomposition. 
While this popular semi-manual approach is commonly used in industry to create  quality hex meshes, existing attempts to replicate the process automatically have failed to produce similar quality results.  We successfully automate this process, producing results on-par with those obtained using manual decomposition on a wide range of inputs.  Our method produces all-hex meshes on the vast majority of inputs (
Figure~\ref{fig:mosaic}), and introduces highly localized non-hex elements when necessitated by the input feature curve networks (Figure~\ref{fig:mixed}). 
As an additional benefit, the user can be optionally involved in the meshing process, customizing the decomposition via a simple user interaction (Figure~\ref{fig:pinion}).

Our method is based on two simple observations. We first note that we can obtain a decomposition that produces well-shaped hexahedral elements by forming a set of {\em cutting surfaces} bounded by {\em cutting loops} distributed strategically across the input model's surface. We further note that by selecting a set of cutting loops that are aligned with a surface cross-field and which interpolate the input features, we can produce a decomposition that respects the geometric characteristic of the input model.

Following this observation, our meshing method computes cutting loops and cutting surfaces bounded by them that satisfy the combination of the criteria above. 
Starting from a cross-field which is aligned to a set of semantic or geometric feature curves (Figure~\ref{fig:teaser}a), we extract a set of well-distributed field-aligned loops on the object surface (Figure~\ref{fig:teaser}b). We facilitate formation of simple low curvature loops, and corresponding well-shaped cutting surfaces, by balancing field alignment against loop geodesicity. 
We generate and dynamically replenish a pool of candidate cutting loops. We repeatedly select the most impactful cutting loops from this pool, ones farthest from previously selected ones.  
We form cutting surfaces that interpolate these loops and adapt to the shape of the input model by using level sets of a smooth volumetric field; we simplify cutting surface geometry by enabling formation of cutting surfaces bounded by multiple loops. 
We use these surfaces to decompose the volume into coarse simple polyhedral blocks, terminating the process once the blocks satisfy our quality requirements (Figure~\ref{fig:teaser}c). We refine the resulting blocks via midpoint subdivision producing a coarse hex-dominant mesh that can then be used to produce high quality hexahedral meshes across a range of resolutions (Figure~\ref{fig:teaser}d).
We validate our framework by testing it on multiple inputs of varying complexity, exhibiting a range of geometric and user prescribed features 
(Section~\ref{sec:results}).  
We were able to generate feature-preserving field-aligned meshes across all tested models, and to obtain all-hex meshes on the great majority of them.
The minimal element quality of the meshes we produced across all models was 0.33, comfortably above the minimal threshold of 0.2 typically expected by industrial users~\cite{Pebay}. We highlight the advantages of our method  by comparing our results to those produced using a range of existing strategies.

 
Our overall contribution is a robust and fully automatic method for block decomposition based hexahedral (or hex-dominant) meshing. We generate comparable or better quality meshes than previous approaches, while guaranteeing feature-preservation and surface-field alignment. 
This contribution is made possible by our use of field-aware surface loop computation, 
generation of valid, smooth cutting surfaces that interpolate these loops, and the loop evaluation procedure that leads to robust detection of most effective loops to embed into the block structure. 
\section{Background and Related Works}
\label{sec:related}

\paragraph{Hexahedral meshing}
The generation of high-quality hexahedral meshes is a well researched open problem, which is regarded as extremely difficult among practitioners~\cite{Shepherd2008,blacker2000meeting,Owen}. 
Quality meshes are expected to satisfy two core requirements:  mesh elements have to
be maximally close to cuboids, and the outer surface of the mesh has to closely approximate the
input model. 
Additional requirements 
include~\cite{blacker2000meeting}:
{\em geometric matching}, which requires meshes to include all input surface feature curves in the mesh edge network; {\em boundary sensitivity}, which implies strong preference for meshes aligned with surface curvature directions; and {\em orientation insensitivity}, which requires the mesh to be independent of the model orientation. 
Unfortunately, 
none of the existing methods can robustly satisfy the union of the five requirements above on general inputs. 

Traditional approaches to hexahedral meshing can be roughly characterized into a few categories~\cite{Shepherd2008,blacker2000meeting}: primitive-based, indirect, advancing-front, grid-based, skeleton-based, and decomposition. 
Primitive based methods and indirect methods
are only capable of producing acceptable quality meshes on narrow sets of geometries~\cite{Shepherd2008}.

Advancing front methods~\cite{tautges1996whisker,Plastering1,Kremer2014} seek to propagate an existing quad surface mesh inward. Recent research on field-based hexahedral meshing can be seen as an extension of this 
approach: the methods first propagate a tensor field from the surface into the interior of the model and then use this volumetric field
to compute a mesh~\cite{Nieser2014,Li:2012,jiang2014frame,kowalski2016smoothness,liu2018singularity,solomon2017boundary}. 
A core advantage of both recent and traditional advancing front approaches is the ability to support all three of the additional requirements above. 
Unfortunately, both types of approaches do not robustly generalize to models with complex topology. 
As noted by Liu et al.~\shortcite{liu2018singularity} existing volumetric field methods require manual intervention to obtain desirable results, even for relatively simple shapes (see Figure 13 in their paper).
Similar to these field-based methods, we use a surface cross-field as a starting point, but instead of propagating it into the model's interior, we use it to guide the formation of cutting loops and surfaces, sidestepping the challenge of processing interior singularities, and achieving all the three goals above. 

Grid~\cite{Lin:2015:QGA:2828638.2828677,Schneiders1996} and octree-based methods~\cite{Marechal2009,ito2009octree} use a regular or adaptive grid to mesh the interior of the input models, and use different 
strategies to connect this mesh to the model's surface. 
While these methods can robustly handle the core hex-meshing requirements, the meshes they produce are orientation dependent and are not aligned with curvature directions. Capturing surface features using these approaches requires excessive local refinement (Figure~\ref{fig:polycube_comparison} middle), and can result in feature loss. 

Polycube-based meshing methods map generic 3D shapes to orthogonal polyhedra (or \emph{polycubes}~\cite{tarini2004polycube}), 
mesh this polycube using a regular grid, and then map the resulting mesh back to the original shape~\cite{Gregson:11, fu2016efficient,huang2014,Livesu:2013:PolyCut,Fang:2016}.
These methods produce grid-connectivity meshes in the interior of the model and can be seen as a generalization of the grid-based approach. 
Similar to grid-based approaches, these methods are orientation dependent and may not match surface features, even after mesh refinement 
(Figure~\ref{fig:polycube_comparison} left).

Automated decomposition techniques seek to cut the object into parts, which can then be meshed conformingly using existing algorithms. Inside-out skeleton \cite{LAPS17,Livesu2016} and medial-axis based decomposition approaches~\cite{li1995hexahedral,Sheffer1999,Quadros} fail to generalize to complex shapes. Methods that start from a dense hexmesh and derive a coarse block decomposition from it~\cite{gao2017simplification,Cherchi,gao2015hexahedral} may fail to align with features that were not present in the input mesh. Surface-driven block-decomposition techniques use surface features to guide cutting surfaces.
Cuts can be used to define both the primal~\cite{cooper,BlackerMulti,RuizGironesMultisweep,Liu97automatichexahedral} or the dual structure~\cite{GAO2018}. Dual methods hardly support feature alignment, because cuts do not directly produce mesh edges, and geometric snapping should be used in post processing to conform with the input features. 
Primal methods are orientation independent, and frequently satisfy both geometry matching and boundary sensitivity. Unfortunately, both the established~\cite{Shepherd2008} and more recent~\cite{WANG2017,kowalski2012fun} automatic block-decomposition methods are limited in the set of geometries they can be applied to.
\begin{wrapfigure}[11]{r}{0.1\textwidth}
\vspace{-1em}
\centering
\includegraphics[width=0.1\textwidth]{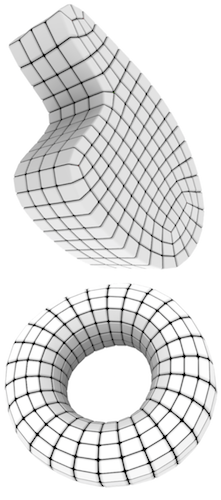}
\end{wrapfigure}
They thus are highly reliant on the sharp feature networks on the model surfaces, and cannot process free-form natural shapes, or shapes with smooth and rounded features. 
Our method follows the block-decomposition approach popularized by these techniques, and inherits their core advantages. However, contrary to those methods, it does not rely just on the feature curve networks on the input models, 
and can robustly handle generic free-form models with both smooth and sharp features (Figure~\ref{fig:hand_joint}). In particular, the torus and the wave in the inset are examples which~\cite{GAO2018} and~\cite{WANG2017}, respectively, list as failure cases.

\begin{figure}
\includegraphics[width=.9\columnwidth]{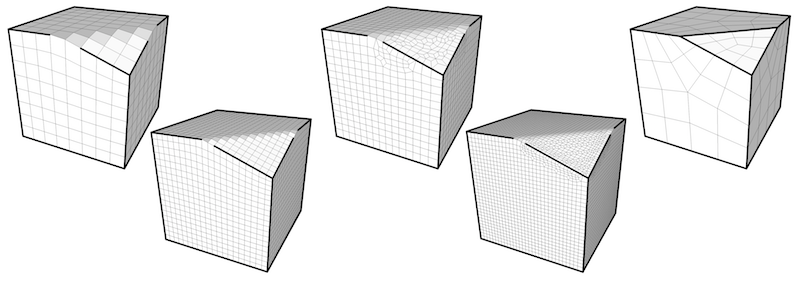}
\caption{Polycube-based methods 
(left), and grid-based methods 
(middle), fail to conform to features that do not align to the major axes, deviating from the surface and introducing elements with non planar facets. While geometric fidelity can be achieved via refinement, features cannot be matched (bottom). Our method aligns to features in any orientation already at a coarse scale, adjusting the singularity structure of the mesh mesh accordingly (right). 
}
\label{fig:polycube_comparison}
\end{figure}

\begin{figure}
\includegraphics[width=\linewidth]{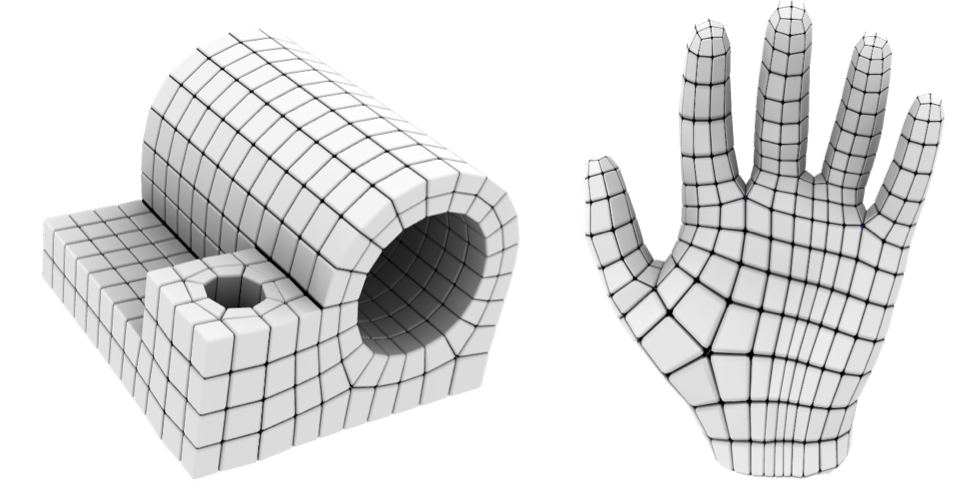}
\caption{State of the art tools based on volumetric fields~\cite{liu2018singularity} require manual intervention to mesh even simple objects like the joint and hand models. Loopy Cuts automatically produces hexahedral meshes with comparable singular structure, without requiring a volumetric field.}
\label{fig:hand_joint}
\end{figure}

\paragraph{Hex-dominant meshes.}
A range of methods~\cite{LevyLP,hybridHexa,ray2017hexahedral,Sokolov:2016:HM:2965650.2930662} support computation of mixed element meshes 
aiming to minimizing the percentage of non-hex elements while satisfying mesh quality requirements. 
While recent hex-dominant methods are very robust in terms of the inputs they handle, 
they still produce a significant percentage of non cuboidal elements \cite{hybridHexa,ray2017hexahedral}.
Our approach produces non-hex elements only incident to singular (non valence-3) surface feature vertices (Section~\ref{sec:results}). 


\paragraph{Cross-fields and field-coherent loops.}
Generating and tracing direction fields on surfaces, or other spatial domains, is becoming a fundamental preprocessing step in numerous applications in computer graphics and geometry processing \cite{vaxman2016directional,de2016vector}. 
Paths traced using most existing methods are not designed to be closed, and are typically terminated when approaching a singularity or another similarly directed path.  A range of recent methods seek to connect cross-field singularities with short, field aligned paths \cite{boiermartin2004ptm,carr2006rectangular,daniels2009semi}. We trace closed field-coherent loops away from singularities following the approaches of \cite{campen2012dual,Pietroni:2016}, which both rely on the formalism introduced in  \cite{kalberer2007qsp}. We use the discrete graph based structure of \cite{Pietroni:2016} to efficiently trace such loops and compute field-aware geodesic distances.

\section{Overview}
\label{sec:method}
The input to our method is a 3D model described by a closed triangle mesh $\SM$,  together with a set of line features demarcated as chains of edges in $\SM$.
We compute a cross-field on the surface $\SM$ using state-of-the-art methodology~\cite{bommes2009mixed,ComplexRoots:Diamanti:2014}. We constrain the field to to follow the line features in the input and the main curvature directions elsewhere. 
We incrementally decompose the model into blocks using a sequence of cuts, each of which cuts through the entire model, producing conforming blocks with shared surfaces. Our choice of cutting surfaces is guaranteed to produce blocks whose boundary edges incorporate the input feature curves and are aligned with the input cross-field. 
The assembly of such blocks, namely their combinatorial structure, will be referred as a \emph{meta-mesh} $\MM$; this structure is updated after each cut and it is formed by cells that correspond to the individual blocks.

Once our incremental cutting results in blocks that are simple enough, we can consider an actual geometrical representation of the meta-mesh $\MM$ composed from polyhedral cells whose faces correspond to portions of the original surface $\SM$ or of the cutting surfaces and whose edges and vertices correspond to the intersections between these different surfaces (note that cell faces need not be planar). 

\subsection{Decomposition Goals}
\label{sub:polyhedra}
%

To obtain the desired output hex (or hex-dominant) mesh quality we aim for 
the meta-mesh to satisfy the following topological and geometrical criteria. We strictly enforce the Boolean criteria when possible and optimize the continuous ones.   
\begin{description}
\item[t1.] Each cell must have genus zero.
\item[t2.] Each cell face must be bounded by at least three edges.
\item[t3.] Each vertex of each cell should have valence three.
\item[g1.] Each cell should be convex.
\item[g2.] Each cell should be well shaped, i.e. have planar faces and orthogonal edges.  
\item[g3.] Each cell should approximate its corresponding  block within a given accuracy.
\end{description}

\begin{wrapfigure}{r}{0.12\textwidth}
  \begin{center} \hspace{-0.04\textwidth}
    \includegraphics[width=0.15\textwidth]{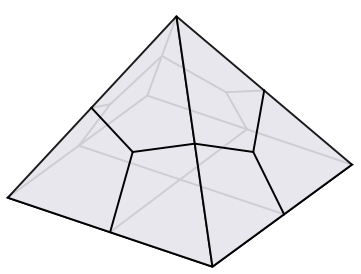}
  \end{center}
\end{wrapfigure}
Criteria \textbf{t1} and \textbf{t2} guarantee that subdividing each cell would produce a polyhedral mesh. Criterion \textbf{t3} guarantees that each cell can be split into hexahedra with a single step of midpoint subdivision. 
However strictly satisfying it while still incorporating all surface line features may in practice be impossible -- see, for instance, the Schneider's four sided pyramid~\cite{Shepherd2008} (inset) -- when the combination of sharp and user drawn feature curves results in an input for which no practical hex-mesh exists. 
We therefore require our method to produce cell vertices of valence three away from irregular feature-curve network vertices, but place no constraints on the valence of cell vertices placed at irregular network vertices. Consequently, absent such vertices our method is guaranteed to produce an all-hex mesh. If such vertices are present the non-hexahedral elements are constrained to the cells containing them. 

Criteria \textbf{g1} and \textbf{g2}  control the quality of the meta-mesh cells and indirectly determine the shape of the elements in the final hex mesh.  As noted by Owen~\shortcite{Owen} a single non-convex element in a hexahedral mesh makes a simulation using this mesh suspect. Finally, criterion \textbf{g3} ensures that the meshes produced by refining the meta mesh and projecting the new vertices to the corresponding block surfaces produces meshes with comparable quality to that of the cells. 
We address such criteria indirectly, as explained in the following subsection.

\subsection{Algorithm}
\label{sub:algoritm}
Informally speaking, we incrementally slice the model by applying cuts, which are defined by loops that strictly adhere to the input line features and follow the input cross field. 
We update the meta-mesh during this incremental process, and stop the process once the decomposition satisfies the above goals. 
Once this decomposition is complete we refine it using mid-point subdivision, converting all cells with valence-three vertices into hexahedra. We then refine the mesh to the user designed density and optimize the mesh quality, using off-the shelf optimization code~\cite{Livesu:2015}. 
%


\noindent\paragraph{Generating ordered loops. Section~\ref{sec:loops}}
In order to obtain an even, field-aware, distribution of the cuts that respect the input line features, we strategically specify the order in which we choose the cut-defining, or cutting, loops.
For this purpose we assemble a preliminary ordered queue of loops that we later scan to generate the cuts that define our decomposition.

\begin{wrapfigure}{r}{0.09\textwidth}
  \begin{center} \hspace{-0.04\textwidth}
    \includegraphics[width=0.10\textwidth]{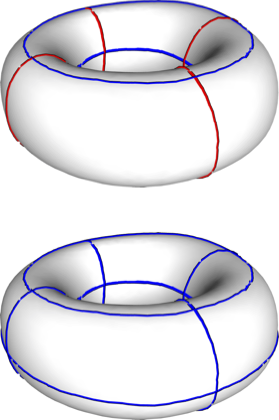}
  \end{center}
\end{wrapfigure}
We determine the order as follows: first, we insert the loops that are needed to interpolate the input features (Sec.~\ref{sub:sharp_feature_trace}); we then add loops that, respecting the cross-field, progressively cover the surface in a uniform way  (Sec.~\ref{sec:loop_distr}). 
We achieve this increasingly denser and denser uniform distribution by defining a distance over the space of loops, and applying a biased furthest-first sampling procedure using this metric. We bias this sampling procedure to satisfy criterion \textbf{t2} by making sure that each loop in the queue is intersected by other loops at three or more points (see inset: the red loops have only two intersections; adding one more loop takes them all to three intersections, giving a configuration sufficient to produce a valid meta-mesh). Thanks to the alignment with the cross field, the selected loops intersect (near)-orthogonally, satisfying criterion \textbf{g2}. 

\paragraph{Processing the loop queue, Section~\ref{sec:cut}}\noindent 
We build our collection of cutting surfaces by repeatedly extracting the top loop from the ordered queue and using it as a starting point for construction of the cutting surface. 

We note that in many instances, such as when forming genus zero blocks starting from a higher genus surface, or when processing models with deep concavities, it may be necessary or simply better to form cutting surfaces bounded by more than one loop (see Figure \ref{fig:cavities}). 
We therefore develop a loop grouping technique that, given a loop extracted from the queue, finds additional loops 
such that the combined set jointly bounds a well formed cutting surface.  
The resulting set of one or more loops partitions the input surface into two disjoint charts. We compute a harmonic field in the interior of the model whose isosurfaces smoothly interpolate these charts. We use the isosurface approximately equidistant from these two charts  as our cutting surface (Sec.~\ref{sec:volcut}). 

Once a surface is computed, we use the intersections between it and the current set of blocks to refine the block structure and the meta-mesh by defining new cells, faces, edges and vertices. 
To satisfy criterion \textbf{t3}, the cutting process constrains new meta-mesh vertices to lie at the intersections of three surfaces, ensuring they have valence three in all their incident cells.
Thus all cells away from singular feature-network vertices in our meta-mesh are split into hexahedra during subdivision.  
Moreover, 
since surfaces intersect nearly orthogonally, we are able to warrant that criterion \textbf{g1} will be satisfied. 

We continue to cut the model using new loops from the queue until all cells satisfy our topological requirements  and provide a good approximation of the outer shape (criterion \textbf{g3}).
The quality of approximation is evaluated simply by measuring the difference of area between each external face of the meta-mesh $\MM$ and its corresponding patch of surface on mesh $\SM$.


\ignore{
********************** OLD PART START HERE ******************
\subsection*{Algorithm (old)}
Given the input described above, our decomposition process is initialized by computing a queue of potential cutting loops, whose ordering is designed to achieve maximally uniform cut distribution across the input surface.   
It then proceeds to decompose the model using a sequence of cuts by picking the top loop from the queue, computing its corresponding cutting surface, embedding it in the current block decomposition and its corresponding meta-mesh representation. 
Throughout the process the method replenishes the queue of potential loops if and when necessary, maximizing a loop-space distance between existing and newly computed loops. 
\enrico{This is not "throughout the process", this happens only during loop selection, not during cut. The latter sentence should be moved before the previous one.}
The process terminates once all cells of the meta-mesh 
satisfy a set of criteria reflecting the goals described above. 
Once the process is complete we refine the decomposition using mid-point subdivision  converting  all cells with valence-three vertices into hexahedra. We then refine the mesh to the user designed density and optimize the mesh quality, using off-the shelf optimization code~\cite{Livesu:2015}. 
An overview of our framework is shown in Figure \ref{fig:teaser}.

\paragraph{Cutting Loops Computation and Ordering, Section~\ref{sec:loops}}
We employ the input cross field to construct cutting loops that balance field alignment against geodesicity. 
We initialize our loop queue by first tracing loops that incorporate 
feature curves (Section~\ref{sub:sharp_feature_trace}) and placing those at the top of the queue.
We grow the queue by sampling additional 
loops along the model surface using a distance metric defined on the space of loops (Section~\ref{sec:loop_distr}). 
This process produces an ordered set of roughly uniformly distributed loops, each maximally away from the previously formed ones, which cut the surface along both cross-field directions.
We achieve this uniform distribution by defining a distance over the space of loops, and applying a furthest-first sampling procedure using this distance metric.
Thanks to the alignment with the field, the selected loops intersect (near)-orthogonally, satisfying criterion \textbf{g2}. 

\begin{wrapfigure}{r}{0.2\textwidth}
  \begin{center}
    \includegraphics[width=0.15\textwidth]{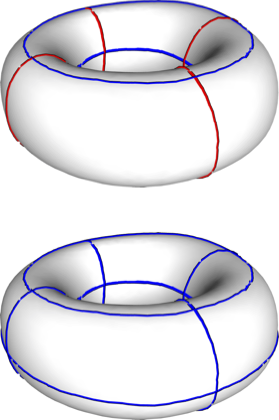}
  \end{center}
\end{wrapfigure}

To satisfy criterion \textbf{t2}  we need to ensure that that each loop in the selected set is intersected in at least three points, thus after adding each loop we add loops that intersect it into the queue \nico{queue??} if the criterion is not yet satisfied (see inset).
We order the queue by giving the highest priority to those loops that fix at least one necessary intersection for a loop previously in the queue, and order the rest based on farthest distance from the previously constructed loops (Section \ref{sub:discretization}). 

\enrico{Probably the paragraph could end here. The rest is technical and explained later.}
\alla{is this still relevant or a repeat?}
Since sampling over the infinite space of loops would be impossible, we discretize the sampling process by generating stochastically a \emph{pool of loops} to sample from .
The pool is initialized with a set of loops that are orthogonal to the given line features, plus an additional set of loops through points evenly distributed on the surface; every time a new loop is selected through sampling, we replenish the pool by adding more loops that are orthogonal to it.   

\alla{This IMHO should go in the loop formation - explaining that we form loops which are pure surface loops and that do not induce cuts}
\paolo{This paragraph is not very clear here. Not sure if it is the right place to discuss convex-concave features} Note that we use all loops in the selected set to generate cuts; the only input features that are retained as outer edges of the meta-mesh without generating any cut are those lines explicitly marked as being convex. 
\alla{do we want to mention here the two loops via a concave feature?}\nico{agree maybe we should remove this.. to early}

\paragraph{Cutting, Section~\ref{sec:cut}}
We form cutting surfaces by repeatedly pulling the top loop in the current queue \nico{check queue} and using it as a starting point for cutting surface construction. 
We note that in many instances, such as forming genus zero blocks starting from a higher genus surface or when processing models with deep concavities, 
it may be optimal to form cutting surfaces bounded by more than one loop (see Figure \ref{fig:cavities}). 
We develop a grouping technique that given a loop finds additional loops, such that the combined set jointly bounds a well formed cutting surface.  
The resulting set of one or more loops is defined so as to partition the input surface into two disjoint charts. We compute the cutting surface that s approximately equidistant from these two chats  by computing a harmonic field in the interior of the model with appropriate boundary conditions set along each chart; the cutting surface is then obtained as an isosurface of such field.  

Once a surface is computed, we use the intersections between it and the current set of block to refine the block structure and the metamesh by defining new cells, 
faces, edges and vertices. 
The cutting process constrains new meta-mesh vertices to lie at the intersections of three surfaces, thus ensuring they have valence three in all their incident cells, which is needed to satisfy criterion \textbf{t3}. 
Thus all cells away from singular feature-network vertices in our meta-mesh are split into hexahedra during subdivision.  
\alla{which fact?  this needs better wording}
This fact, combined with the fact that surfaces intersect nearly orthogonally, warrants that also criterion \textbf{g1} will be satisfied. 

We continue to cut the model using new loops, until all cells satisfy our topological requirements  and provide a good approximation of the outer shape (criterion \textbf{g3}) \alla{here add reference or details for how we measure this one}
If the queue is depleted at any point in the process we replentish it using the same strategy as above.

}


\begin{figure*}
\begin{tabular}{@{}c@{}c@{}ccc@{}}
\begin{tabular}{@{}c@{}}
\includegraphics[height=0.18\linewidth]{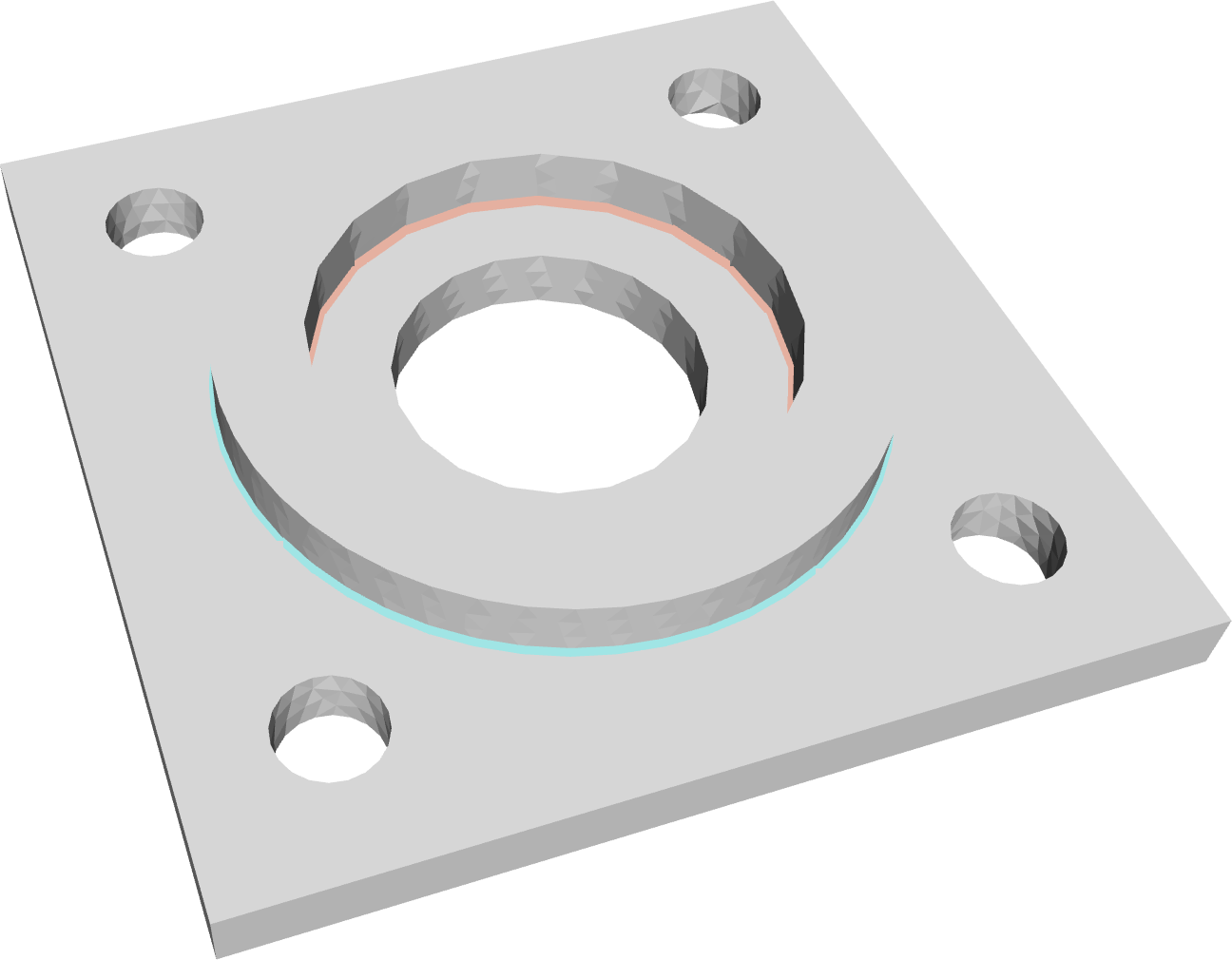}
\end{tabular}&
\begin{tabular}{@{}c@{}}
\includegraphics[height=0.18\linewidth]{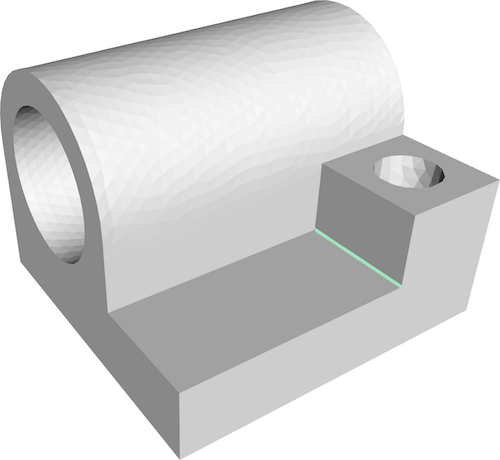}
\end{tabular}&
\begin{tabular}{@{}c@{}}
\includegraphics[height=0.10\linewidth]{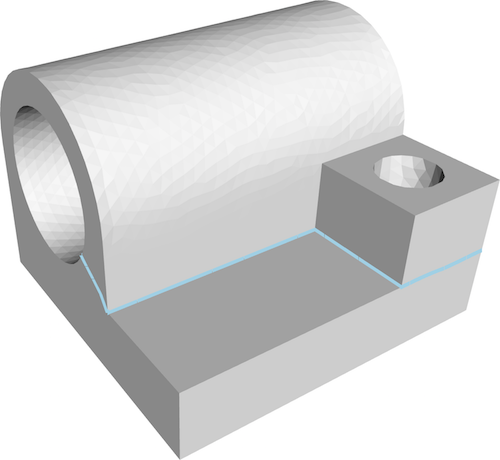} \\
\includegraphics[height=0.10\linewidth]{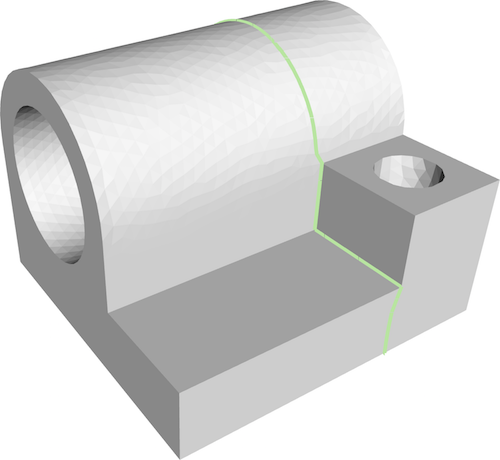} 
\end{tabular}&
\begin{tabular}{@{}c@{}}
\includegraphics[height=0.18\linewidth]{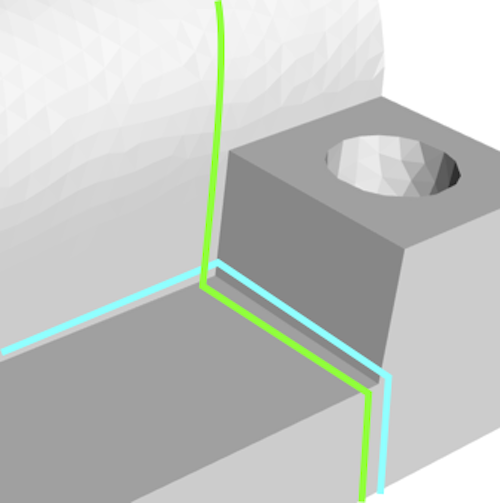}
\end{tabular}&
\begin{tabular}{@{}c@{}c@{}}
	\begin{tabular}{@{}c@{}}
		\includegraphics[height=0.08\linewidth]{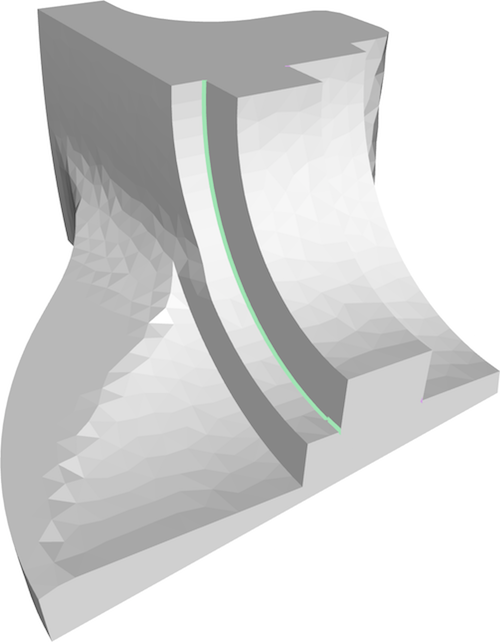}\\
		\includegraphics[height=0.08\linewidth]{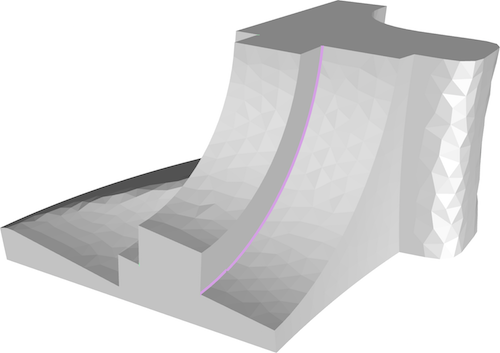}
	\end{tabular}&
	\begin{tabular}{@{}c@{}}
		\includegraphics[height=0.2\linewidth]{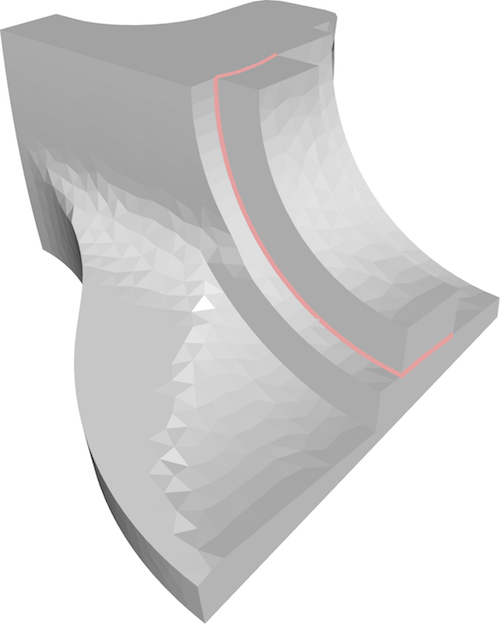}
	\end{tabular}
\end{tabular}\\
(a)&(b)&(c)&(d)&(e)\\
\end{tabular}
\caption{(a) Closed concave curve features; (b) an open curve feature; (c) the open curve feature is extended to form loops; (d) a closeup on field topology of traced loops (displaced from the feature in the rendering); (e) a concave loop connecting multiple concave incomplete sharp features (left, the open features; right, the loop connecting them).}
\label{fig:sharp_feature_completion}
\end{figure*}

\section{Computing Cutting Loops}
\label{sec:loops}

Given a cross field $\X$ over a surface $\SM$, we trace field-coherent geodesic paths w.r.t.\ $\X$, as in \cite{Pietroni:2016}.
A formal definition of field-coherent paths and loops rests upon the stratification $\SMF$ of $\SM$ as defined in \cite{kalberer2007qsp} and  briefly summarized in Appendix~\ref{app:prelim}.
Informally we define a \emph{field-coherent geodesic loop} through a point $p\in\SM$ to be a closed curve that goes through $p$, is loosely following one of the directions of $\X$ and is as short as possible according to the anisotropic distance defined in Equation \ref{eq:drift} (Appendix~\ref{app:prelim}).  
Roughly speaking, field-coherency forces a loop to approximately follow the underlying direction field 
until it gets back to its origin.
Requiring a curve to follow $\X$ exactly is unlikely to produce closed loops, as tracing the filed directions exactly would usually result in a spiral effect; allowing the curve to drift from $\X$ while remaining as short as possible allows us to use $\X$ as a guide to extract consistent loops.

Each point $p$ on $\SM$ which is away from a field singularity can be crossed by exactly two field-coherent geodesic loops, which are  orthogonal to one another at $p$ (disregarding the traced curve's orientation). We consider the set $\LL$ representing the space of all loops for any point $p\in\SM$. 

We incrementally build a queue of loops $\L \subset \LL$, which will be scanned to generate the cuts, as follows:
\begin{enumerate}
\item We initialize $\L$ with loops that incorporate a subset of the input curve features  (Sec.~\ref{sub:sharp_feature_trace});
\item We extend $\L$ using a sampling process that aims to select loops that intersect orthogonally and are evenly distributed on $\SM$. This step is based on furthest sampling on $\LL$ and requires the definition of a suitable distance on the space of loops 
(Sec.~\ref{sec:loop_distr}).
\end{enumerate} 
Since the space of loops $\LL$ is infinite, the sampling process is discretized by forming an extensive \emph{pool of loops} $\Pool$ from which we sample, and which is built dynamically during the selection process 
(Section~\ref{sub:discretization}). 

\subsection{Generating Loops Incorporating Line Features}
\label{sub:sharp_feature_trace}
The extraction and classification of sharp input features is out of the scope of this paper and they are considered part of the input. 
Our initial feature set consists of both open and closed curves. We form a feature curve network by placing vertices at feature curve intersections, sharp corners, and dangling end-vertices of open curves. We break the curves into segments bounded by vertices and treat each segment as a separate curve in the process below.   
We classify each feature curve as \emph{concave}, \emph{convex} or \emph{flat} based on the average dihedral angle along it.  
We employ this classification to determine the number of cutting loops we want to incorporate each feature curve into: we expect each flat feature to belong to a single cutting loop and expect each concave feature to belong to two cutting loops.  We do not perform cuts along the convex features, but use them to initialize the meta-mesh (Section~\ref{sec:init_meta}). This cutting loop formation strategy fosters the formation of as-right-as possible dihedral angles for all blocks incident to a given feature (Figure~\ref{fig:concave_cuts}). 
 


We initialize the cutting loop queue $\L$ as follows:
\begin{enumerate}
\item We add to $\L$  all loops formed by closed non-convex features.
\item We 
extend each open concave feature into two complete loops by loop tracing and extend each open flat feature into one loop (Appendix \ref{app:prelim}). 
\item For each convex feature that has one dangling endpoint we try to extend it into a loop or, if not possible, we trace a loop through that endpoint in the direction orthogonal to the feature curve, adding it to the queue. This step helps eliminate vertices of valence 1 or 2 on the subsequent meta-mesh which would induce non-hexahedral blocks.
\end{enumerate}
Figure \ref{fig:sharp_feature_completion} shows some examples of concave line features and their extension to form loops. 
Note that
loops should not pass through singularities (see Appendix \ref{app:prelim}), while the endpoints of an open feature $\Lf$ are likely to lie at singularities. 
We get around this issue by tracing two loops $\Li{1}$ and $\Li{2}$ that run parallel to the feature $\Lf$ along its two sides, and infinitesimally close to it.
The portions of the loops $\Li{1}$ and $\Li{2}$ corresponding to the feature $\Lf$ are constrained to follow it. After completion these portions are snapped to $\Lf$, while the remaining portions are left free to follow the field 
elsewhere.
Since the endpoints of $\Lf$ might be singularities, the loops $\Li{1}$ and $\Li{2}$ may take different routes (see Figure \ref{fig:sharp_feature_completion}.c-d). 
We favor the formation of loops that encompass more than one line feature (see Figure \ref{fig:sharp_feature_completion}.e),
by introducing a bias during loop tracing, which reduces the length of curves running along curve features (see Section \ref{sub:discretization} and Appendix~\ref{app:concave-tracing} for details). 

We discard self intersecting loops. When tracing loops we
prevent any new loop from tangentially intersecting, or partially overlapping with, another loop or feature curve (other than the one it lies on). See Appendix \ref{app:prelim} for a formal definition of tangential and orthogonal intersections. To this end we form a set of constrained edges $\CF$ initialized with all convex feature curves. For any newly formed loop we prevent it from tangentially intersecting curves in $\CF$ or $\L$. 

\subsection{Sampling Loops}
\label{sec:loop_distr}

We augment the queue $\L$ with additional loops in order to obtain a fairly regular distributions of loops over $\SM$ and to allow for formation of blocks that satisfy the topological constraints in Section~\ref{sub:polyhedra}.  We define an \emph{arc}  as a portion of a loop $\Lo\in\L$ that lies between two consecutive intersections of $\Lo$ with other loops of $\L$.
To satisfy criterion \textbf{t2} we require each loop to consist of at least three arcs.

%
 


\subsubsection{Loop Space Distance}
In order to evenly sample loops from the space $\LL$, we introduce a notion of (non-symmetric) distance between loops, as the average over one loop of the shortest distance from each of its points to the other loop. 
Let $\Li{i}$ and $\Li{j}$ be two loops, then we define:
\begin{equation}
d(\ell_{i},\ell_{j}) = \frac{1}{|\ell_{j}|}\int_{\ell_{i}} \mbox{dist}(\ell_{i},\ptheta) d\ptheta
\label{eq:dist}
\end{equation}
where $\mbox{dist}(\ell_{i},\ptheta)$ is the length of the shortest field-coherent geodesic path joining a point of $\Li{i}$ to $\ptheta$.
In order to get an intuition for this distance, consider that two roughly parallel loops on $\SM$ are close to one another, while loops that are either intersecting orthogonally, or wind about different handles of an object (of non-null genus) are usually far apart. 
 
Now considering a set of loops $\mathcal{L}=\{\ell_{1},\ldots,\ell_{k}\}$, and a loop $\Lo$ not belonging to $\L$, we can generalize 
\begin{equation}
d(\mathcal{L},\ell) =    \frac{1}{|\ell_{j}|}\int_{\ell_{i}}  \mbox{min}_{\ell_i\in\mathcal{L}}\mbox{dist}(\ell_{i},\ptheta) d\ptheta,
\label{eq:distset}
\end{equation}
hence the notion of farthest loop $\bar{\ell}$ from $\L$ is well-defined as \[\bar{\ell}=\mbox{argmax}_{\ell\in\mathbf{L}}d(\mathcal{L},\ell).\]

\subsubsection{Loop Insertion}

We extend the queue $\L$ incrementally through a process of farthest point sampling. 
We form a discretization $\Pool$ of the infinite space of loops $\LL$,  which we refer to as the {\em pool}, which is built dynamically and used to sample loops to build $\L$ as follows. We define the set $\overline{\L}=\L\cup\CF$ as the set that contains all loops in $\L$ together with the set of all convex curve features.
We form the initial pool as follows.
\begin{enumerate}
\item We sample all curves of $\overline{\L}$ at fixed intervals and, for each sample, we trace orthogonal loops that we add to $\Pool$. Traced loops are constrained to avoid tangential intersections with the elements of $\overline{\L}$;
\item We perform a Poisson point sampling on $\SM$ \cite{Corsini2012} to obtain a set of seeds $\PSeed$ and, for each seed $p$, we trace the two orthogonal loops, each constrained by the elements of $\overline{\L}$ as above, and we add such loops to $\Pool$. 
\end{enumerate}
Each time we add a new loop $\Lo$ to $\L$, we replenish the pool with new loops obtained by sampling $\Lo$ as described in item (1).  
When adding loops to $\Pool$, we prevent them from tangentially intersecting any loop in $\L$. 


We add loops from the pool to the queue $\L$ using the following priority rule.
Let $\hat{\L}$ be the subset of $\L$ made of loops that are formed by less than 3 arcs. 
We give higher priority to those loops in $\Pool$ that intersect at least one loop in $\hat{\L}$; among them, we select the loop $\Lo$ that maximizes its distance from all loops in $\overline{\L}$ and we add $\Lo$ to $\L$. 
Next, 
we remove from $\Pool$ all loops that intersect $\Lo$ tangentially, and retrace them from their sources in $\PSeed$, constrained to the updated $\overline{\L}$. 

\begin{figure}
\begin{tabular}{@{}cccc@{}}
\includegraphics[width=0.22\linewidth]{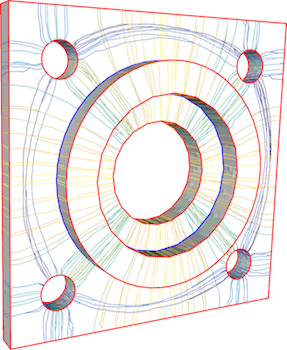} &
\includegraphics[width=0.22\linewidth]{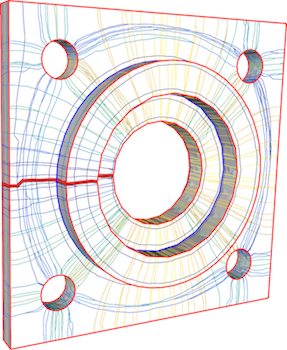} &
\includegraphics[width=0.22\linewidth]{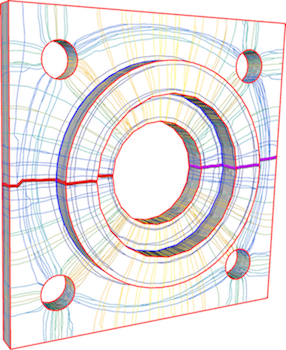} &
\includegraphics[width=0.22\linewidth]{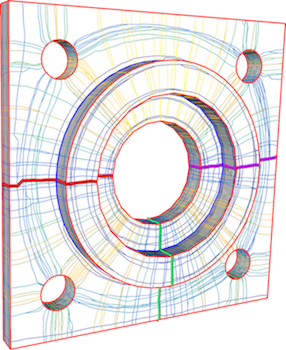}\\
\includegraphics[width=0.22\linewidth]{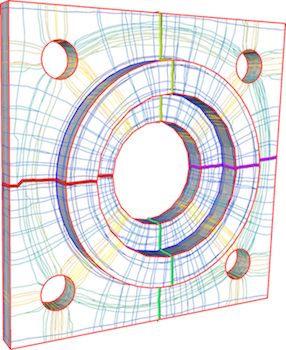} &
\includegraphics[width=0.22\linewidth]{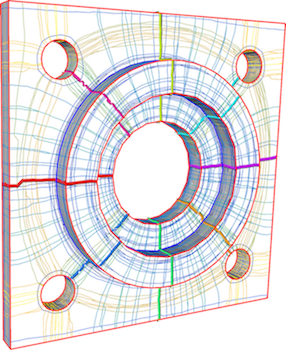} &
\includegraphics[width=0.22\linewidth]{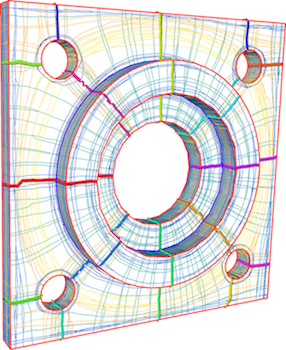} &
\includegraphics[width=0.22\linewidth]{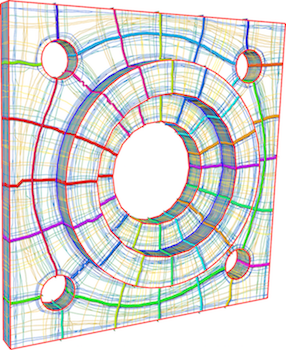}\\
\end{tabular}
\caption{The queue of loops $\mathcal{L}$ is created iteratively (red lines). The thin lines denote loops in the pool $\mathcal{P}$, where color denote distance from the features already present in $\mathcal{L}$: blue close, yellow far.}
\label{fig:loop_distribution}
\end{figure}

This step can be repeated at will and fosters the formation of a queue $\L$ that contains loops with at least three arcs and are uniformly distributed over $\SM$. 
Figure  \ref{fig:loop_distribution} shows a few steps of loop generation (see Section \ref{sub:discretization} for details). Note in Figure \ref{fig:loop_distribution} how both the collection of loops in $\L$ and the pool $\Pool$ become progressively more dense as the process goes on.
The generation of loops is stopped as criteria \textbf{t2} and \textbf{g3} are both satisfied (where geometric fidelity \textbf{g3} is tested on the patches intercepted by the network of loops on mesh $\SM$); in order to speedup the process, criterion \textbf{g3} is not tested until \textbf{t2} is satisfied at all loops.

\subsection{Discrete Loop Tracing}
\label{sub:discretization}
To efficiently and robustly implement the tracing procedure and furthest sampling process described in the two previous sections, we trace field-coherent geodesic loops  with the discrete graph-based approach of \cite{Pietroni:2016}, which brings to the stratified structure $\SMF$ (Appendix \ref{app:prelim}) the technique of \cite{Lanthier:2001} to evaluate geodesic paths and distances. 
In short, in \cite{Lanthier:2001} shortest paths and distances are found by a Dijkstra search on an extended graph $\G$, which is built over $\SM$ edges and vertices, plus Steiner points sampled on edges and arcs connecting vertices of $\SM$ and Steiner points across each triangle. 
In our case, four point-arrows are generated per vertex and per Steiner point, which are properly arranged on $\SMF$, and just field-coherent arcs connecting them are considered. 
The graph $\G$ is built once, and used in all subsequent processing. 
One important advantage of this method is that crossings and overlaps of paths can be handled in a robust, combinatorial way that does not involve numerics: 
it is possible to precisely identify whether two paths intersect orthogonally or tangentially by simply comparing arcs that belong to the same triangle of $\SM$ and checking their underlying direction fields.  Thus during tracing we prevent any new loops from tangentially intersecting loops in $\L$ by blocking Dijkstra propagation along arcs in $\G$ that are orthogonal to those arcs belonging to paths already in $\overline{\L}$. 

\paragraph{Loop Space Distance}
Given a queue of loops $\L$, we set a source for each node of graph $\G$ traversed by each loop $\Li{i}\in\L$ and we run a Dijkstra propagation.
The distance $d(\L,\Lo)$ from any other loop $\Lo$ is computed easily by collecting distances at all nodes traversed by $\Lo$ after propagation. 
Note that a single Dijkstra propagation is sufficient to set distances at all nodes of $\G$, thus it is sufficient to evaluate distances from all loops. 

\paragraph{Extending curve features to loops}
At the beginning of the process, all open line features must be extended to loops, as described in Section \ref{sub:sharp_feature_trace}.
Since most such features join singularities of $\X$, we cannot directly extend them from their endpoints, which are not represented in graph $\G$. 
Therefore, we proceed by tracing loops that run parallel to each feature, and we snap such loops to the features later on.
This procedure is described in more details in Appendix~\ref{app:concave-tracing}.

\begin{figure}
\begin{tabular}{@{}c@{}c@{}}
\includegraphics[width=0.47\linewidth]{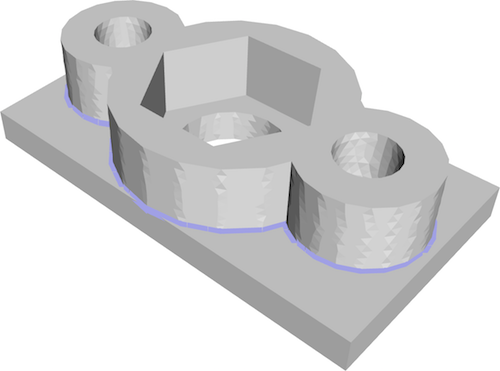}& 
\includegraphics[width=0.47\linewidth]{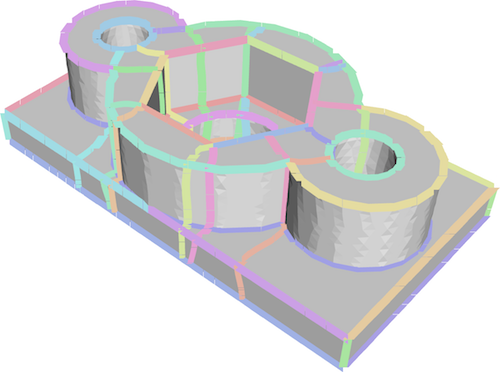} \\
(a)&(b)
\end{tabular}
\caption{(a) a loop spanning across multiple concave features; (b) a mesh where all sharp features have been sampled.
}
\label{fig:multiple_concave_features}
\end{figure}

A complete set of sampled sharp features for a mesh is shown in figure  \ref{fig:multiple_concave_features},b.  
A sequence showing the first three loops sampled for a simple cube is shown in figure \ref{fig:loop_distribution}. 


\subsection{Convex Features and Meta-Mesh Initialization}
\label{sec:init_meta}

We initialize the meta-mesh $\MM$ as being formed of a unique cell, with surface $\SM$, embedding on it all convex features as edges, and their corners and endpoints as vertices; faces of $\MM$ are the components of $\SM$ that are disconnected by this network of convex features (possibly, there can be a single face with cuts). 
For each convex feature that has one dangling endpoint  we extend the feature along its direction to form a loop. We add the edges along this loop to the meta-mesh, updating the face structure accordingly. 

\section{Model Cutting}
\label{sec:cut}
Our cutting method receives as input the current top loop $\Li{i}$ in the loop priority queue, and uses it to form a cutting surface that partitions the current model $\SM$ into two disjoint parts, refining the intersected blocks and their corresponding meta-mesh cells. 
Each cutting surface is bounded by a set of loops that includes $\Li{i}$.
The cutting process involves three major steps: loop grouping,  cutting surface computation, and block set and meta-mesh refinement.

\subsection{Loop grouping}
\begin{figure}
\includegraphics[width=\linewidth]{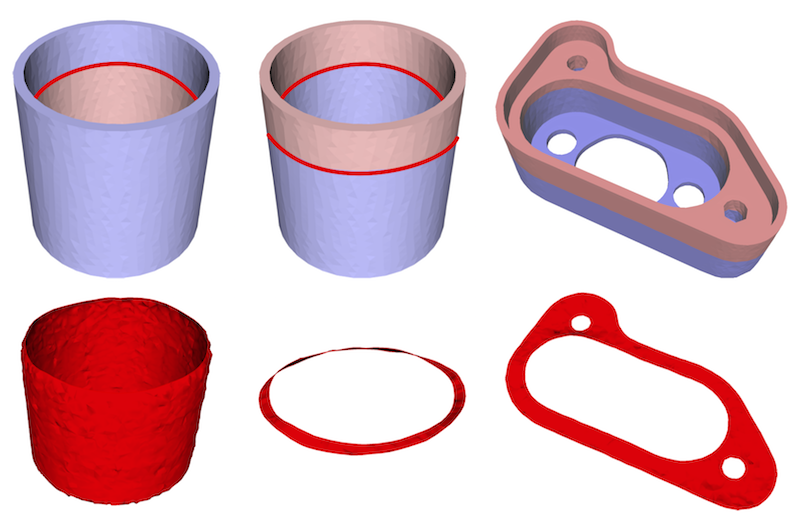}
\caption{Left: halving a genus zero object using a loop located inside a cavity generates a highly non-planar cutting surface (red). Middle: matching the inner loop (type II) with its best matching loop outside the same cavity (type I)  produces a better cut. Right: pairing one external loop with multiple inner loops we provide high quality surface cuts for complex shapes containing a variety of tunnels and cavities.}
\label{fig:cavities}
\end{figure}

While a cut bounded by a single loop is enough to halve a genus zero object, additional loops may be necessary to cut in half higher genus shapes.
Moreover, using a single loop 
may produce highly non-planar cuts in the presence of deep cavities. Using surfaces with multiple boundary loops in such cases 
allows for cuts with less geometric distortion (Figure~\ref{fig:cavities}).
We address both scenarios by forming cutting surfaces with multiple boundary loops, producing better shaped blocks. We group loops together using a combination of topological and geometric criteria.

\begin{figure}[h]
\includegraphics[width=\columnwidth]{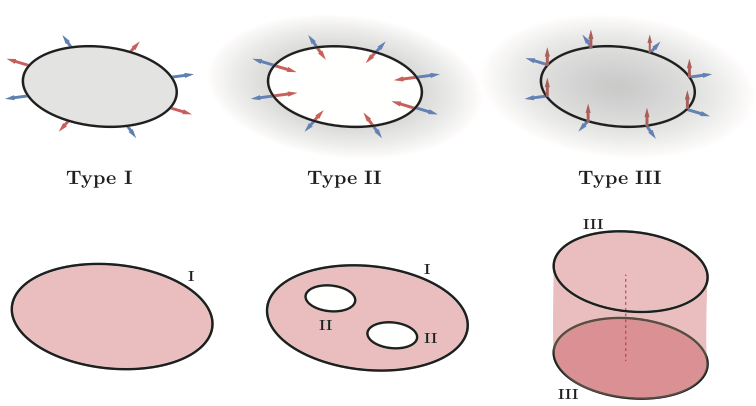}
\caption{Top: loops are grouped according to the type of cut they induce on the object, which can be inferred by studying the relation between the outgoing loop normals (blue) and the surface normals (red).  Bottom: loops whose normals are roughly orthogonal to the surrounding surface (type I and II) can be combined together to split high genus objects. In particular, type II loops arise in tunnels and deep cavities, and always require a pairing loop of type I for cutting. Loops whose normals lie in the surface's tangent plane (type III) require a complementary loop for cutting and can be matched only with loops of the same type.}
\label{fig:loop_types}
\end{figure}

We observe that, from a cutting perspective, we need to distinguish between three different types of loops, which can be roughly classified according to the interaction between the outward loop curve normals
and their coincident surface normals (Figure~\ref{fig:loop_types}):
\begin{itemize}
\item \textbf{Type I} loops conceptually lie outside the object, and have outward normals well aligned with the 
surface normal. These loops may bound a well shaped, or near planar, cutting surface on their own, and form the majority of our cutting surface boundaries.   
\item \textbf{Type II} loops have normals that point in the opposite direction to the surface normals, and are typically located inside tunnels or cavities. To obtain a well shaped cutting surface, these loops always require a cutting mate of type I, which defines the outer boundary of the same cutting surface, and may occasionally be grouped with other loops of same type, generating a surface with multiple holes.
\item \textbf{Type III} loops have normals that lie close to the surface tangent plane. To provide a non-degenerate partition of the object, they  need to be paired with a cutting mate of the same type, to jointly form an approximately cylindrical cutting surface.
\end{itemize}
Note that this characterization is not formal, but rather a tool to predict the shape of the surfaces that would result from using each loop or a set of loops as a boundary of a cutting surface. 
In fact, a cut that interpolates two loops of type III and a cut that interpolates one loop of type I and one loop of type II are topologically equivalent (they are both generalized cylinders with two disjoint boundaries); but while in the former case the cylinder has a well defined inner axis, in the latter case it may be completely foreshortened with two coplanar boundaries.

\paragraph{Loop assessment}
To assign a unique type to each loop we study the relation between its normals and the surface normals (Figure~\ref{fig:loop_types}). We compute loop normals using the curve's Frenet frame, oriented according to the winding order that provides normals that largely point away from the loop's center of mass. 
We integrate the dot product between loop and surface normals, and perform the classification by applying a symmetric threshold centered at 0 to discriminate between the three types. Values higher than 0.3 denote type I loops; values lower than -0.3 denote type II loops; and values in between denote type III loops.

\begin{figure}[ht]
\includegraphics[width=\columnwidth]{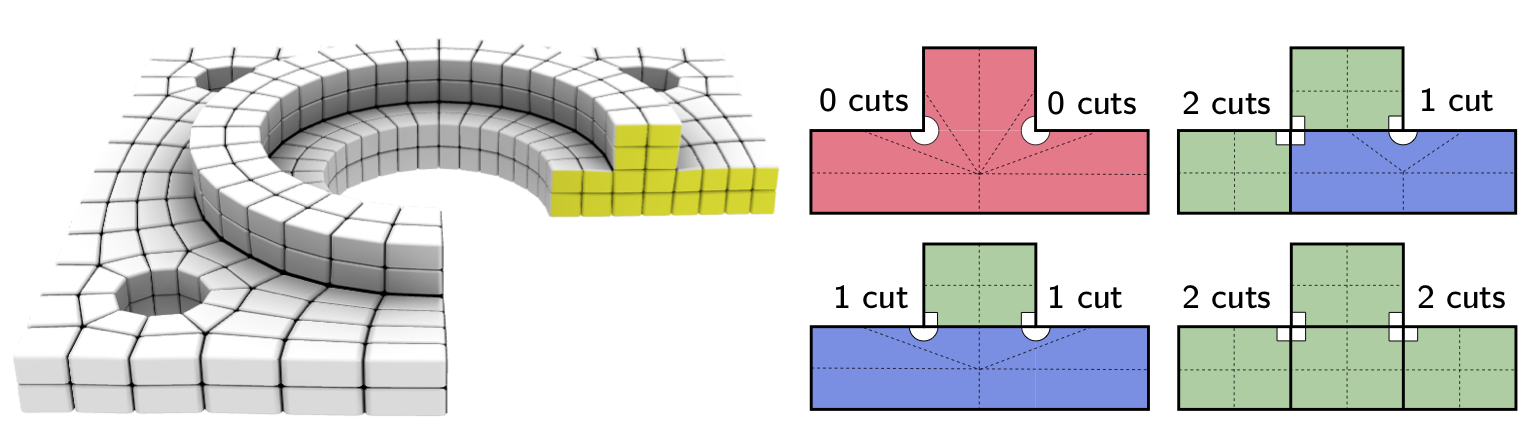}
\caption{Concave features have an internal angle larger than $\pi$, and always require two orthogonal cutting surfaces to produce a high quality block decomposition. Note that the two cuts at each concavity interpolate loops of different types: the horizontal cut interpolates two loops of type I and II; the vertical cuts interpolate two loops of type III each. Thin dashed lines show how the decomposition would look like after one step of midpoint subdivision.}
\label{fig:concave_cuts}
\end{figure}

Note that if a loop spans a sharp geometric feature there will be a mismatch between the surface normals on one side of the loop and the normals on the opposite side. As explained in Appendix~\ref{app:concave-tracing}, our loop tracing method always produces two loops 
along each concave feature on its opposite side. Each loop is infinitesimally close to the feature, and is parallel to it without ever crossing it. Consequently each such loop has a different surface normal and induces different cuts. At the same time, in geometric space the portions of the loops along the feature completely overlap. Thus along closed features such as the two concave rings of the Bearing Plate, the loops generate two cuts orthogonal to one another emanating from the same geometric loop (Figure~\ref{fig:concave_cuts}).

\paragraph{Grouping}
Starting from a loop $\Li{i}$, we wish to find the smallest set of complementary cutting loops that jointly with this loop cut the surface in half and produce a cutting surface whose shape is most reflective of the loop's type. If the loop $\Li{i}$ is of type I or II, we seek to obtain a cutting surface that is as planar as possible, whereas if the loop $\Li{i}$ is of type III we wish to obtain a cylinder-like cut.  We reflect this preference using a loop pairing metric defined as follows:
$$
E(\ell_{i},\ell_{j}) = \left\lbrace
\begin{array}{ll}
E_\mathsf{plane}(\ell_{i},\ell_{j}) & \text{if } \ell_{i},\ell_{j} \text{ have both type I or II}\\
E_\mathsf{cyl}(\ell_{i},\ell_{j}) & \text{if } \ell_{i},\ell_{j} \text{ have both type III}\\
\infty & \text{otherwise}\\
\end{array}\right.
$$
where the terms $E_\mathsf{plane}$ and $E_\mathsf{cyl}$ measure the similarity between the loops spanning planes or cylinders, respectively. 
Specifically, $E_\mathsf{plane}$ is $\epsilon$ if $\ell_{i}$ and $\ell_{j}$ span the same approximating plane, and grows proportionally to the angular and euclidean distance between the planes they span. $E_\mathsf{cyl}$ is $\epsilon$ if $\ell_{i}$ and $\ell_{j}$ span the same approximating cylinder, and grows proportionally to the angular distance between their axes and the ratio between their cylinder radii. Both metrics are defined in the range $(0,1]$ and implemented using Gaussian functions. The details on the computation of the approximating planes and cylinders and the metric used to assess their similarity are provided in Appendix~\ref{app:pairing}.

\begin{figure}
\includegraphics[width=\linewidth]{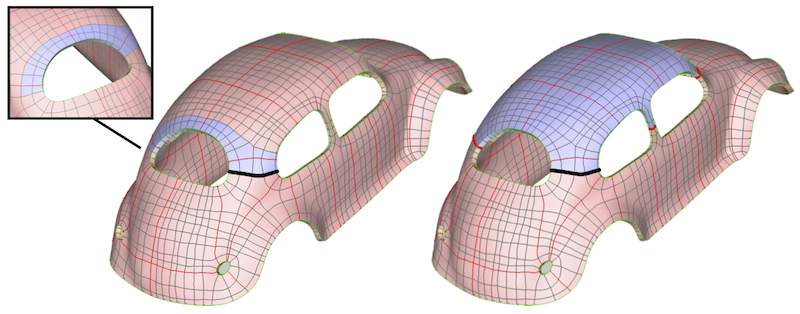}
\caption{Left: Using a loop that does not have a good geometric matching with any other loop in the queue may produce globally inconsistent bi-partitions, where only a portion of the edges participating in a loop are selected for cutting (closeup, bottom left corner). Right: when the geometric grouping fails, we use topological grouping as a backup strategy. This strategy minimizes cut length and therefore performs minimal cuts around topological handles (right).}
\label{fig:pairing}
\end{figure}

Given an initial loop $\Li{i}$, we determine the optimal set of cutting loops by computing a binary labeling of the surface $\SM$ that minimizes
\[
E(\ell_{i}) = \sum_{\ell_j \in \mathcal{L}} E(\ell_i, \ell_j)\:,
\]
and is constrained to have different labels along the opposite sides of $\Li{i}$ to guarantee that the boundaries of the bi-partition include it. We compute the solution using a min-cut formulation applied to the dual of mesh $\SM$. 
Cuts along mesh edges (arcs in the dual graph) that do not belong to any loop receive $\infty$ costs, whereas a cut along a mesh edge $e$ that belongs to some loop $\Li{j_e}$ contributes to the energy proportionally to the ratio between its length and the total length of the loop it belongs to
\[
	E_e(\Li{i},\Li{j_e}) = E(\Li{i},\Li{j_e}) \cdot \frac{\vert e \vert}{\vert \Li{j_e} \vert}
\]
The minimizer of $E(\Li{i})$ automatically provides the smallest set of additional loops that globally halve $\SM$ and are geometrically best aligned with $\Li{i}$. 
We compute the solution using the min-cut implementation provided by~\cite{boykov2001fast,kolmogorov2004energy,boykov2004experimental}.
Note that minimizing $	E$ at local (per edge) level may produce inconsistent global results, where only a subset of the edges participating in some loop is selected for cutting (Figure~\ref{fig:pairing}). Empirical observations indicate that this happens only if there are no good geometric matches for $\Li{i}$ in the loop queue $\L$.
We recover from these pathological situations by using a backup strategy, which consists in minimizing the total length of the cuts, as detailed in Appendix~\ref{app:pairing_backup}. 
The backup strategy is purely topological, and essentially cuts along  all topological handles that are necessary to globally halve the shape with as short as possible cuts. 
Since loops are geodesic, 
cutting along a whole loop in general produces a shorter boundary than partially cutting along multiple loops, and therefore a valid cut set is usually found. 
If no valid cutting loop set can be found also with the backup strategy, the initial loop is discarded. 

\paragraph{Handling Cavities}
The loop grouping as described above introduces complementary loops in the cut set only if $\Li{i}$ does not bi-partition the surface of $\SM$ alone. If $\Li{i}$ is a separating loop no complementary loops will be found, as they would increase the energy.
As a result, loops of type II located inside cavities will not receive a matching loop of type I, producing highly non planar cutting surfaces that induce a poor block decomposition (Figure~\ref{fig:cavities}, left). 
Thus we first locate all type II loops in our queue and apply the grouping process from them to other loops. If one of those loops is subsequently pulled from the queue we use its previously computed group as the cutting surface boundary.
To provide a quality decomposition of complex objects containing a variety of tunnels and cavities, several inner loops of Type II may need to match to the same outer loop of Type I at once, producing a single surface cut with many holes inside (Figure~\ref{fig:cavities}, right). We obtain the desired effect by imposing specific boundary conditions on the min-cut algorithm. For each loop of Type II, we first compute its best geometric match of Type I using $E_\mathsf{plane}$. We then cluster together all loops of Type II that match the same Type I outer loop, and each time any of these loops is used for cutting, we force the min-cut step to cut through all of them.
Similarly, for each loop of Type III that does not surround a handle (i.e. it is contractible), we find its best geometric matching according to $E_\mathsf{cyl}$, and force min-cut to include both such loops in the bi-partition. 

\begin{figure}
\centering
\includegraphics[width=\columnwidth]{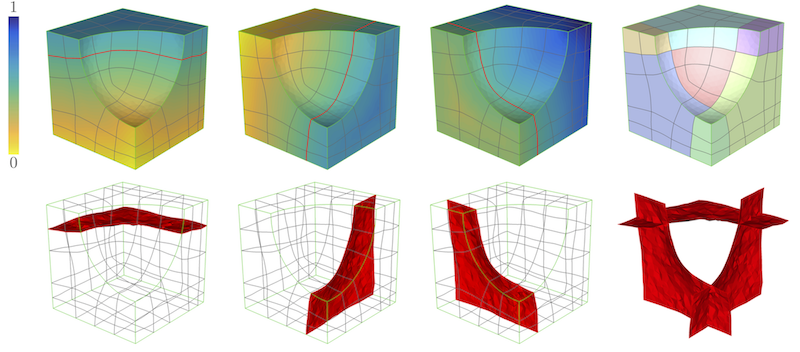}
\caption{Three solid cuts computed with our algorithm. Cuts are defined as level sets of a harmonic function that evaluates to $0.5$ along the cutting loop, is higher on one side of the cut, and lower on the opposite side. Cuts are embedded into the connectivity of the mesh, producing a labeling that induces a solid decomposition of the object (right).}
\label{fig:cuts}
\end{figure}

\begin{figure*}
\centering
\includegraphics[width=\linewidth]{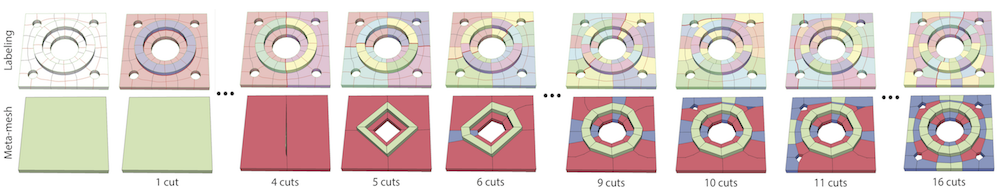}
\caption{Progressive generation of the meta mesh for the Bearing Plate. Top: the labeled tetrahedral mesh; bottom: its corresponding meta-mesh (hexa in green, prisms in blue, other cells in red).}
\label{fig:cut_sequence}
\end{figure*}

\subsection{Solid cut}
\label{sec:volcut}
Given the set of cutting loops $C = \left\lbrace \ell_0, \ell_1,..., \ell_n \right\rbrace$ that bi-partitions the surface of the object, we extend the cut to the interior by defining it as the level set of a volumetric harmonic function that aligns with $C$.
Harmonic functions 
satisfy the maximum principle, which ensures that if the function is non flat, then maxima and minima arise only at the boundaries of the domain and never in the interior. 
By prescribing Dirichlet boundary conditions on the surface $\SM$, we can therefore control the topology of the cutting surface, and guarantee that it interpolates all the cutting loops in $C$, is manifold, and does not interpolate any other points on the model's surface.

Considering the bi-partition of the surface $\SM$ induced by the loop set $C$, we denote $V_C$ as the vertices of these loops, $V_{l}$ as the vertices to the left of $C$, and $V_r$ as the vertices to the right of $C$. We then solve for the volumetric harmonic function $\Delta f = 0$, subject to the following Dirichlet boundary conditions
\[
f = \left\lbrace
\begin{array}{cl}
0.5 & \text{ if } v \in V_C\\
0.5 + \nicefrac{d(v,C)}{2D} & \text{ if } v \in V_l\\
0.5 - \nicefrac{d(v,C)}{2D} & \text{ if } v \in V_r\\
\end{array}
\right.
\]
where the 0.5 level set of $f$ defines the cutting surface bounded by our set of loops $C$ (Figure~\ref{fig:cuts}); and $d(v,C)$ denotes the distance between each surface vertex $v$ and its closest loop in the cut set $C$, measured using a Dijkstra search on the graph of edges of the volumetric mesh $\VM$. The normalization factor $D$ is used to bound the function to the range $[0,1]$, and is the maximum distance from a surface vertex to the set of loops $C$. 
We implement the $\Delta$ operator as a simple combinatorial Laplacian, using uniform weights for all vertices in the one ring.
The choice of combinatorial Laplacian was dictated by the need for solution robustness. Our cutting surfaces are embedded into the connectivity of the tetrahedral mesh and cut across mesh edges. This embedding process progressively worsens the shape of the tetrahedral mesh elements as more cuts are introduced. 
While the popular cotangent Laplacian is geometry-aware and produces smoother level sets on well shaped meshes, 
our experiment show that using it on poor quality meshes results in ill-conditioned and numerically unstable linear systems. 
In contrast, the combinatorial Laplacian is robust to meshing defects, and always yields a well-conditioned positive semi-definite matrix. While in theory it is not geometry-aware, in practice the cutting surfaces we obtained were always good enough for our purposes, even when the tetrahedral elements were poorly shaped.

\subsection{Meta-mesh update}
The meta-mesh $\MM$ is initialized as a single volumetric cell, which incorporates on its boundary all the input feature lines as edges, their endpoints as vertices, and the surface patches of $\SM$ that are disconnected by them as faces. 
Solid cuts are progressively embedded into the connectivity of a tetmesh by splitting all the edges traversed by the 0.5 level set (Figure \ref{fig:cut_sequence}). 
As a result, each cut is a collection of triangular facets of $\VM$ that separates pairs of tetrahedra incident at them at both sides. Using these facets as boundaries and performing exhaustive region growing on the tetmesh we produce and maintain a labeling of tetrahedra defining the volumetric cells of our decomposition. 
We extract the faces, edges and vertices of the associated meta-mesh from this labeling. Each face of the tetmesh at the interface between two different labels (including the null label conventionally assigned to outer space) is labeled with such pair of labels, and all tetmesh faces with the same pair of labels are gathered to form a meta-mesh face; then edges of the meta-mesh are obtained by gathering chains of edges of the tetmesh that have the same set of two or more incident faces of the meta-mesh. 
Finally, tetmesh vertices incident on three or more  edges of the meta-mesh are classified as meta-mesh vertices.
In case topological inconsistencies arise while intersecting a cutting surface with the current meta-mesh, we fix them as explained in Appendix \ref{app:topoclean}.

\begin{figure*}[p]
\includegraphics[width=\linewidth]{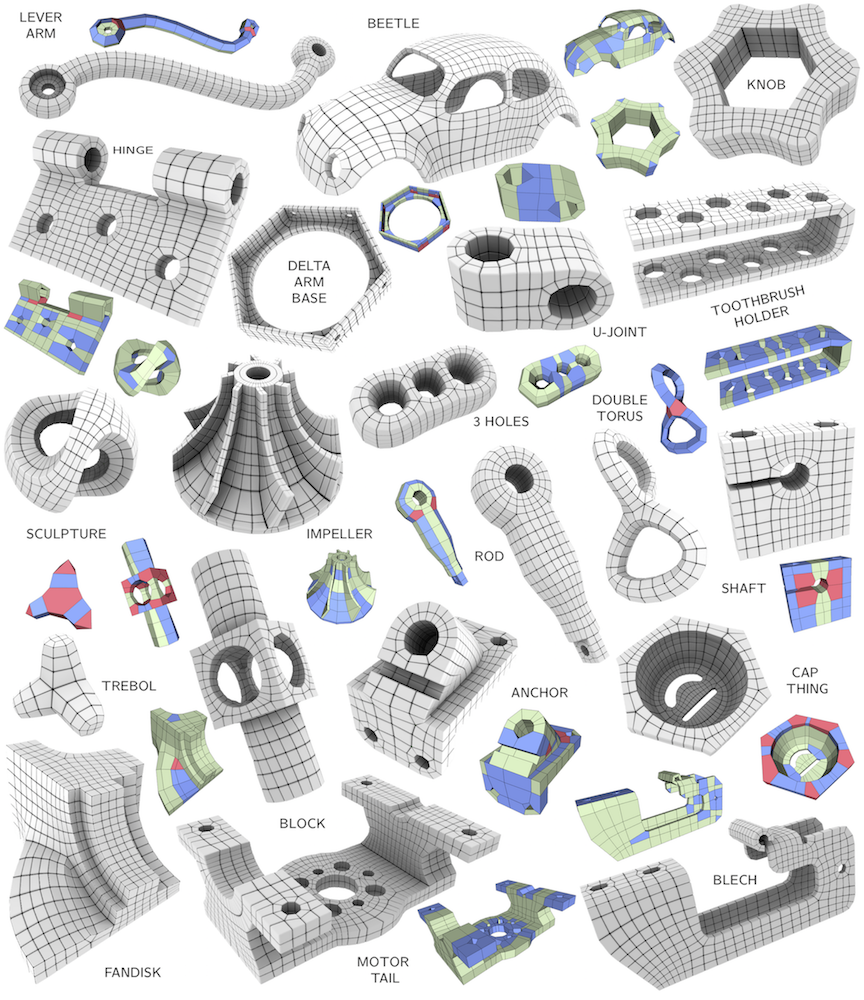}
\caption{Hexahedral meshes produced by Loopy Cuts in automatic mode. Small insets showcase meta-meshes (hexa in green, prisms in blue, other cells in red).}
\label{fig:mosaic}
\end{figure*}

\begin{figure*}[h]
\includegraphics[width=\linewidth]{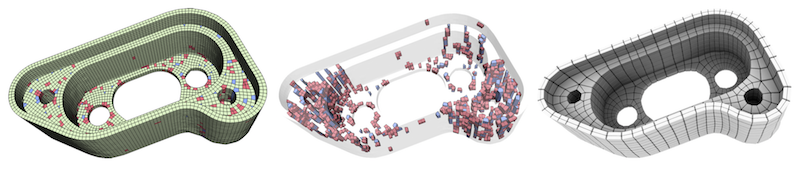}
\caption{Left: a hex-dominant mesh produced with~\cite{hybridHexa} (hexahedra in green, prisms in blue, other polyhedra in red). Middle: the mesh contains many non cuboidal elements, both in the interior and along the exterior. Non hexahedral elements can be arbitrarily complex: in this model the most complex element counts 17 faces. Right: our result.}
\label{fig:lock}
\end{figure*}

\begin{figure}[h]
\includegraphics[width=\linewidth]{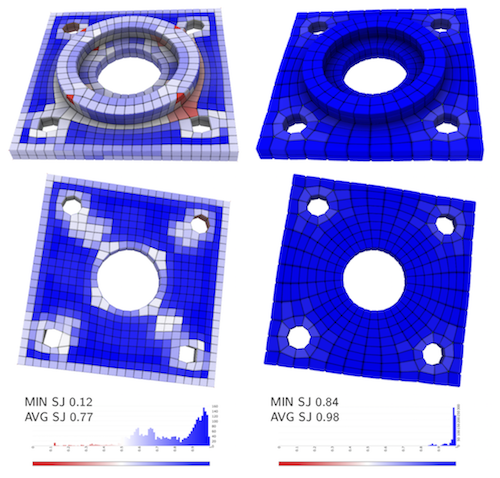}
\caption{Comparison between a state-of-the-art PolyCube-based method~\cite{Livesu:2013:PolyCut} (left) and Loopy Cuts (right). Even though the PolyCut mesh is almost four times larger than ours (2.7K vs 0.7K elements), it is not as effective at capturing the sharp features on the object (see the square-like little holes at the bottom). Meshes derived from our block-decomposition naturally align with features, and yield a quality mesh with much higher average and minimum scaled Jacobians.}
\label{fig:bearing_plate}
\end{figure}

\begin{figure}[h]
\includegraphics[width=\linewidth]{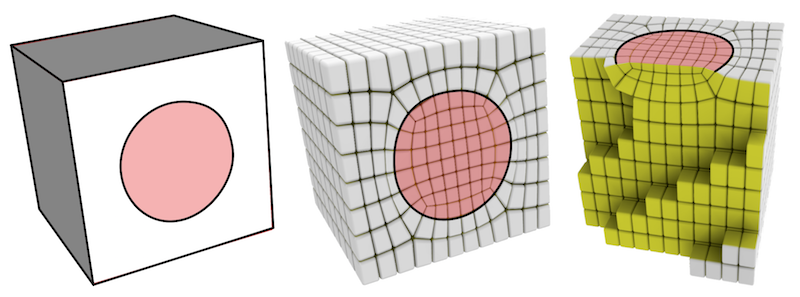}
\caption{Left: an input cube with a user-demarcated circular feature on one of its faces. Middle: the hexahedral mesh produced by Loopy Cuts. Right: cut through view showing the inner mesh connectivity (the cube is rotated to highlight the  singular structure beneath the circular feature).}
\label{fig:cube_w_circle}
\end{figure}

\begin{figure}
\includegraphics[width=\columnwidth]{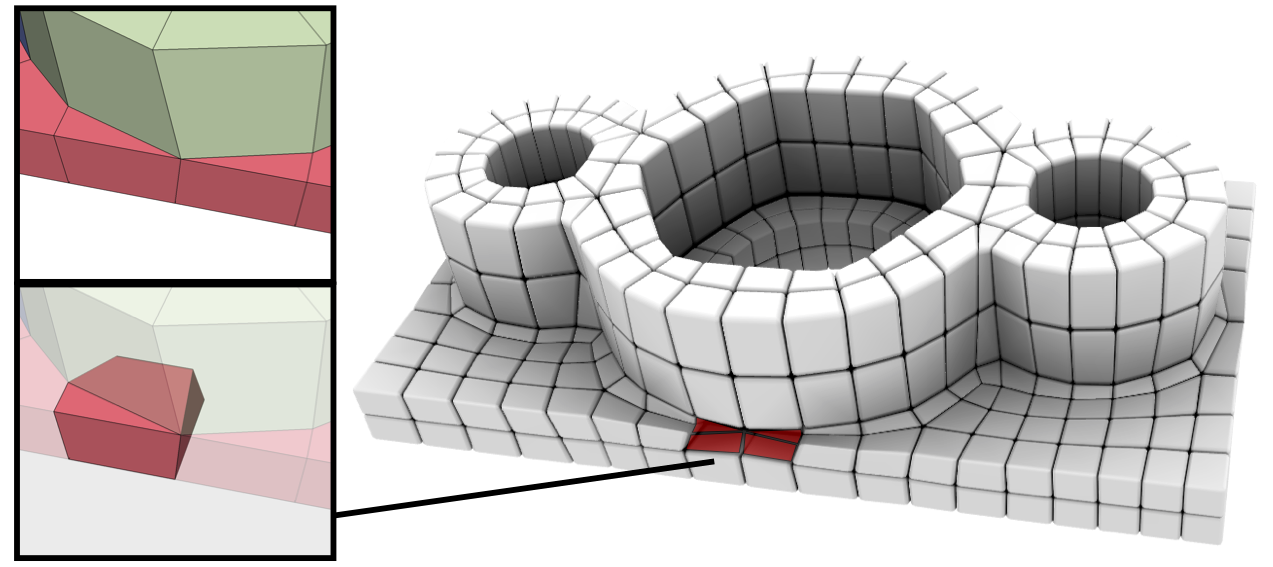}
\caption{Left: Loopy Cuts may introduce mixed elements if surface features that are not cut through form non-valence-three vertices with respect to a meta-mesh cell. Right: the resulting mesh after midpoint subdivision (mixed elements are in red).}
\label{fig:mixed}
\end{figure}

\section{Results}
\label{sec:results}
We validate our method on a range of models of different complexity, both mechanical and organic, and showcase 31 fully hexahedral meshes and just one hex-dominant mesh throughout this paper.  We report numerical statistics in Table~\ref{tab:results}. Loopy Cuts produces meshes with a complex singularity structure, enabling them to conform to the input features and at the same time provide extremely high per-element quality, often much higher than alternative hexmeshing techniques (e.g. Figure~\ref{fig:bearing_plate}). We demonstrate our ability to conform to a variety of features, both geometric (Figure~\ref{fig:mosaic}) and synthetic (Figure~\ref{fig:cube_w_circle}). 
Our pipeline creates pure hexahedral meshes given surface feature networks with all valence three vertices. 
Our midpoint subdivision introduces non hexahedral elements only on meta-mesh cells that are incident at vertices with a valence different from three 
(Figure~\ref{fig:mixed}). 
These configurations seldom occur in practice, and in our experiments we produced only one hex-dominant mesh containing four hybrid elements (less than 1\% of total elements count). As shown in Figure~\ref{fig:lock}, we generate full hexahedral meshes even when state of the art hex-dominant meshing techniques introduce a significant amount of hybrid elements. In general, these techniques consistently produce meshes with a much lower percentage of regular elements (72\% to 91\% hexahedra for~\cite{hybridHexa}, and 33\% to 95\% hexahedra for \cite{Sokolov:2016:HM:2965650.2930662}), and contain complex hybrid elements that may have up to 40 facets (see Table 1 in~\cite{hybridHexa}).

\begin{figure}
\includegraphics[width=\linewidth]{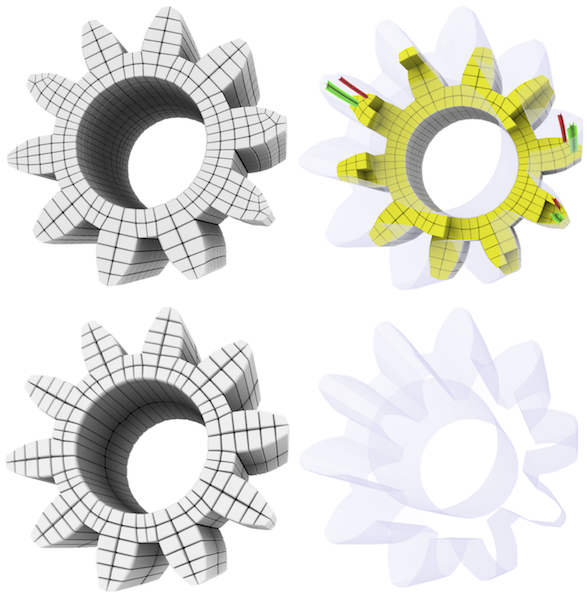}
\caption{Top: in fully automatic mode the cut set and loop grouping used by Loopy Cuts produce a result where different dents have different mesh connectivity, and a few singular chains of edges appear in the interior. Bottom: by manually selecting the cut set (from the automatically generated sequence) and performing guided grouping we produce a perfectly symmetric mesh with no singularities; instead high anisotropy is introduced along the height of the wheel (no horizontal cuts were performed, the only visible cut is due to midpoint subdivision).}
\label{fig:pinion}
\end{figure}

\begin{figure}
\includegraphics[width=\columnwidth]{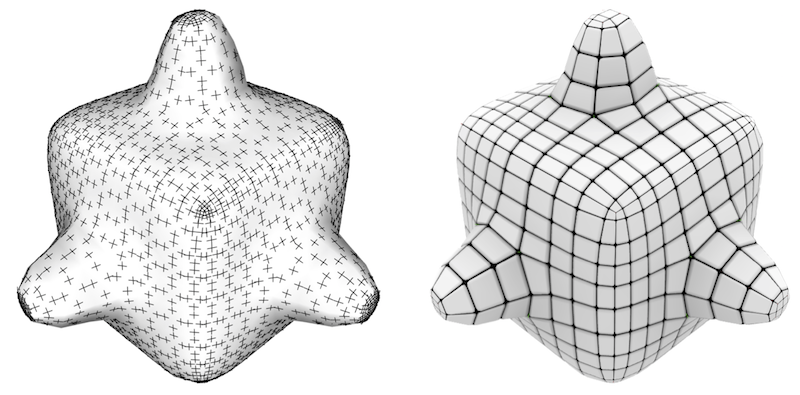}
\caption{The volumetric meshes generated by Loopy Cuts are guided by a surface field. Leveraging existing field processing tools we can therefore control the meshing process. Here we encoded in the cross field the symmetries of the object (from \cite{Panozzo:2012:sym}) that were then transferred to the hexahedral mesh.}
\label{fig:symmetry}
\end{figure}

\paragraph{Customization} Loopy Cuts is fully automatic, but can optionally be used in interactive mode to allow the user to customize the block decomposition. In Figure~\ref{fig:pinion} we show a result obtained with manual interaction, where we guided the loop grouping and cut sequence to obtain a perfectly symmetric mesh with high anisotropy along one direction. We provide this control by enabling three interactive operations that can be activated via a trivial point-and-click user interface. The user can:
\begin{itemize}
\item Prescribe a customized cut sequence. Each time a loop is selected for cutting, the system automatically finds complementary loops, and the volumetric cut is performed;
\item Manually pair loops for cutting. Note that this could also be use to accumulate multiple cuts into one (e.g. the cuts that separate the dents from the pinion wheel in Figure~\ref{fig:pinion} were made all together, while the automatic approach would have processed them in sequence);
\item Mark a loop as a new \emph{convex feature}. Such a loop will then be incorporated in the meta mesh as a chain of surface edges that split faces, but the meta mesh will not be cut through it. 
\end{itemize}
An additional indirect form of customization can be obtained by editing the guiding surface field with user-interactive tools such as the one proposed in Instant Meshes~\cite{jakob2015instant}.  Such tools can be used to prescribe additional soft constraints on the meshing process promoting alignment to secondary features or symmetries (Figure~\ref{fig:symmetry}).


\paragraph{Implementation details.}
We implemented \emph{Loopy Cuts} 
as a single threaded C++ application, using 
Tetgen~\cite{si2015tetgen} for tetrahedralization, and Eigen~\cite{eigenweb} for numerics. Cross fields aligned to line features and surface curvature were computed using MIQ \cite{bommes2009mixed} and the PolyVector Fields method \cite{ComplexRoots:Diamanti:2014}. 
Sharp creases were automatically detected by thresholding dihedral angles, whereas other features were manually marked. 
Note that both field computation and feature detection are external to the algorithm; alternative techniques may also be used, as \emph{Loopy Cuts} is completely agnostic to how such information are computed.  We produce our output meshes by applying one step of midpoint subdivision~\cite{li1995hexahedral} to each block of the meta mesh, and optimizing the geometry with edge-cone rectification~\cite{Livesu:2015}.
Our pipeline runs in minutes; the global running time depends directly on the number of cuts, and indirectly on the complexity of the shape. 
On average, the generation of the cutting loops takes between 10 and 120 seconds on meshes in the range of 4K-50K tris (often obtained by feature-preserving remeshing from high-resolution meshes); while 3 to 8 seconds per cut were necessary to cut models between 100K and 200K tets.

\begin{table}
\small
\centering
\begin{tabular}{| l | r | c | c c c | r c |}
\hline
\textbf{Model} & $\bigtriangleup$ &\textbf{Cuts} & \multicolumn{3}{|c|}{\textbf{Meta-mesh}} & \multicolumn{2}{|c|}{\textbf{Output mesh}}\\
          & 			& 			& 		 \textbf{H} & \textbf{P} & \textbf{O} & \textbf{H}  & \textbf{SJ} \\
\hline
3holes & 19k & 13 & 64 & 16 & 0 & 1K  &   .50/.87\\
anchor & 11k & 16 & 44 & 28 & 9 & 0.7K    & .33/.85\\
anti backlash nut$^*$ & 13k &34 & 77 & 14 & 16 & 0.7K & .43/.84\\
bamboo pen & 36k &20 & 76 & 38 & 0 & 0.8K    & .72/.93\\
bearing plate & 10k & 16 & 43 & 12 & 16 & 0.7    & .84/.98\\
beetle & 70k & 34 & 154 & 38 & 0 & 1.6K    & .50/.95\\
blech & 60k & 42 & 132 & 30 & 0 & 1.3K    & .64/.96\\
block & 28k & 15 & 33 & 32 & 10 & 0.6K    & .41/.82\\
cap thing & 57k & 15   & 68 & 18 & 6 & 0.8K    & .56/.92\\
cube minus sphere &4k & 3  & 17 & 4 & 1 & 0.2K    & .62/.97\\
cube 2 colors & 2k & 11  & 56 & 16 & 8 & 0.8K    & .50/.96\\
spiky cube &6k & 22  & 136 & 72 & 56 & 1.8K    & .57/.93\\
delta arm base &37k & 22  & 52 & 38 & 12 & 0.9K    & .50/.77\\
double torus &32k & 11  & 0 & 88 & 4 & 0.6K    & .49/.81\\
fandisk &7k & 17  & 126 & 11 & 1 & 2K    & .48/.85\\
hand &20k & 29  & 31 & 82 & 37 & 1K    & .33/.83\\
hinge &19k & 13  & 56 & 14 & 2 & 0. 6K    & .52/.94\\
impeller &24k & 31  & 96 & 96 & 0 & 1.7K    & .36/.88\\
joint &9k & 12 & 56 & 12 & 0 & 0.6K    & .77/.97\\
knob &14k & 10 & 88 & 12 & 0 & 0.7K    & .82/.97\\
lever arm &9k & 23  & 36 & 32 & 4 & 1.2K    & .54/.92\\
lock &42k & 22  & 124 & 58 & 18 & 1.7K    & .38/.85\\
motor tail &15k & 45  & 177 & 53 & 0 & 2K    & .51/.95\\
pinion &43k & 40  & 230 & 30 & 0 & 2K    & .41/.95\\
pinion (manual) &43k & 12  & 40 & 0 & 0 & 0.3K    & .66/.95\\
rod &9k & 15 &  108 & 46 & 4 & 1.2K    & .72/.95\\
sculpture &32k & 25  & 80 & 8 & 0 & 0.7K    & .77/.97\\
shaft &30k & 10  & 20 & 32 & 8 & 0.6K    & .44/.92\\
torus &4k & 12  & 65 & 52 & 0 & 0.8K    & .82/.93\\
toothbrush holder &13k & 26  & 43 & 38 & 0 & 0.7K    & .62/.93\\
trebol &13k & 10  & 0 & 15 & 17 & 0.2K    & .47/.77\\
wave &3k & 11  & 40 & 16 & 16 & 0.5K    & .60/.93\\
\hline
\end{tabular}
\caption{Statistics of our method. For each dataset we report: the number of faces of the input mesh $\bigtriangleup$ , the number of cuts being performed; the number of hexa (\textbf{H}), prisms (\textbf{P}) and other elements (\textbf{O}) in the meta mesh; the number of hexa (\textbf{H}) in the final mesh; minimum and average Scaled Jacobian (\textbf{SJ}) of the hexahedral elements of the mesh. ($*\:$ this model is mixed, and contains 4 non-hexa elements).}
\label{tab:results}
\end{table}


\section{Conclusions}
\label{sec:conc}

We present a new approach to block-decomposition for hexahedral meshing. The method combines tracing of strategically placed cross-field coherent cutting surface loops with computation of fair cutting surfaces interpolating one or more such loops. As shown by our results the method outperforms prior work in providing a combination of robustness, feature interpolation, and curvature alignment. 

\paragraph{Limitations and Future work}
Our framework is not guaranteed to compute a valid decomposition. In particular we envision the following sources of failure. First, our loop formation strategy relies on the underlying cross-field. 
On surfaces where the cross-field directions change multiple times, the resulting loops may be too complicated or the tracing may not be able to close loops properly, avoiding 
self-intersections. 
Second, our cutting surface computation depends on the existence of suitable groups of loops that jointly bound a a desired cutting surface. 
Sometimes, some loops find no mate to pair with, just because such loop cannot be found on the sole basis of the existing line features and the cross field. 
Finally, our grouping strategy is heuristic and may fail even when suitable loops exist. 
We did not exploit user interaction at its full potential, since in our current implementation this is allowed only during the pairing and cutting phase; 
enabling simple user interaction during loop selection may resolve with a few clicks most of the issues described above, for instance by forcing or deviating some loop, or changing the flow of the cross-field. 
Addressing all the above aspects is an interesting avenue for future research.

\begin{acks}
This work is partially supported by the EU ERC Advanced Grant CHANGE, grant agreement No.694515.
\end{acks}

\bibliographystyle{ACM-Reference-Format}
\bibliography{biblio}


\appendix
\section{Computing cutting loops}
\subsection{Field-coherent loops}
\label{app:prelim}
\begin{wrapfigure}{r}{0.11\textwidth}
  \begin{center}
    \includegraphics[width=0.1\textwidth]{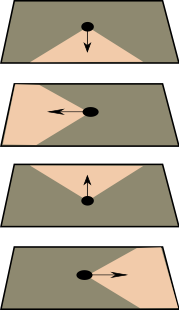}
  \end{center}
\end{wrapfigure}

Following \cite{kalberer2007qsp}, the four components of a cross field $\X$ can be separated on a stratification $\SMF$ of manifold $\SM$ into four sheets, defined as follows (inset). 
For every point $p$ of $\SM$, except the singularities of $\X$, consider four copies $\pz$, $\ppt$, $\pp$ and $\ptp$, each consisting of $p$ together with one of the four directions of $\X$ at $p$, such that $\pz = -\pp$ and $\ppt = -\ptp$.
We call each such copy $\ptheta$ of $p$ a \emph{point-arrow} meaning that it incorporates both a position on $\SM$ and a direction on its tangent space.  
Space $\SMF$ consists of four sheets, each corresponding to $\SM$ less the singularities of $\X$, such that the point-arrows $\ptheta$ defined above are distributed among the layers to form a smooth direction field (see inset). 
Generally speaking, if $\X$ has singularities, the direction field on $\SMF$ turns about such singularities, thus sliding between different sheets, and $\SMF$ consists of a single connected component. 
Space $\SMT$ is the quotient space of $\SMF$ obtained by identifying pairs of point-arrows $\ptheta$ and $-\ptheta$, thus consisting of two sheets, each endowing a line field.  
Manifold $\SM$ can be also seen as a quotient space of $\SMF$, by identifying the four point-arrows at each point $p$.

Following  \cite{Pietroni:2016}, a smooth (oriented) curve $\Lo$ on $\SMF$ is said to be a \emph{field-coherent} path if its tangent direction at all points does not differ for more than $\pi/4$ from the underlying direction field on $\SMF$ (pink wedges in the inset). 
With abuse of notation, we denote by $\Lo$ also the corresponding curves on $\SMT$ and $\SM$, regarded as quotient spaces of $\SMF$.
Two field-coherent paths $\Li{1}$ and $\Li{2}$ are said to intersect \emph{tangentially} if they intersect in $\SMT$; while they intersect \emph{orthogonally} if they intersect in $\SM$ but they do not intersect in $\SMT$. 

The drift of a field-coherent path w.r.t.\ $\X$ comes from the angle between the direction field and the tangent of $\Lo$ at each point along it. 
Following \cite{Pietroni:2016} we adopt an anisotropic metric on $\SMF$ that increases the length of a path proportionally to its amount of drift:
\begin{equation}
\|w\|_{\mathbf{X}}=|w|  (1+\alpha \frac{\measuredangle(\ptheta,w)}{\pi/4}) 
\label{eq:drift}
\end{equation}
where $w$ is a tangent vector at $p$, $|w|$ is its Euclidean norm, $\ptheta$ is the reference direction on $\SMF$, $\measuredangle$ measures the unsigned angle between a pair of vectors, and $\alpha$ is a parameter that tunes the amount of penalty for the drift.  
A field-coherent \emph{geodesic} path between to point-arrows on $\SMF$ is a field-coherent path joining them that is shortest according to the above metric. 
We define a \emph{field-coherent geodesic loop} 
to be a non-null field-coherent geodesic path that starts and ends at the same point. 

\begin{figure}[t]
\begin{tabular}{cc}
\includegraphics[width=0.44\linewidth]{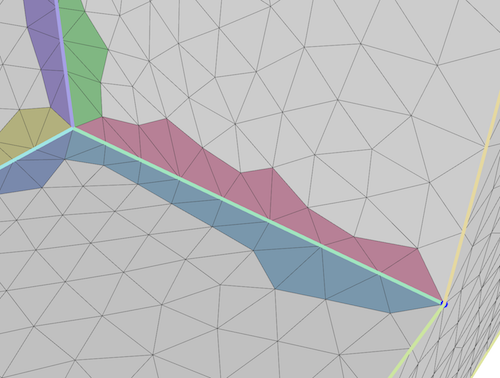} &
\includegraphics[width=0.44\linewidth]{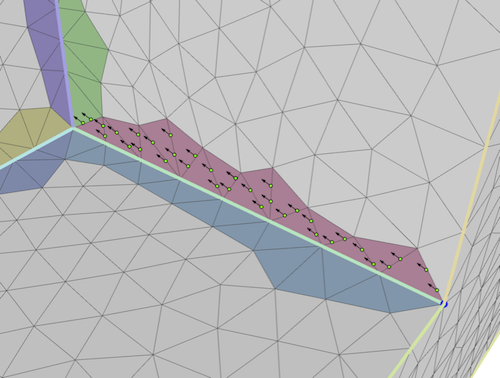}\\
(a)&(b)\\
\includegraphics[width=0.44\linewidth]{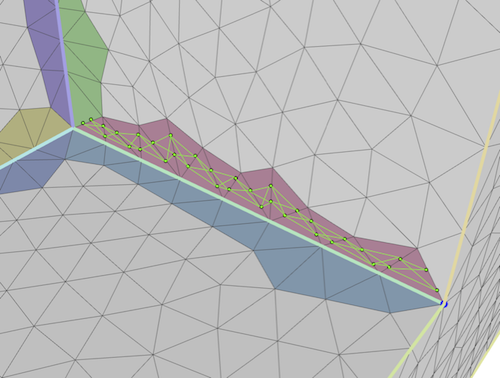} &
\includegraphics[width=0.44\linewidth]{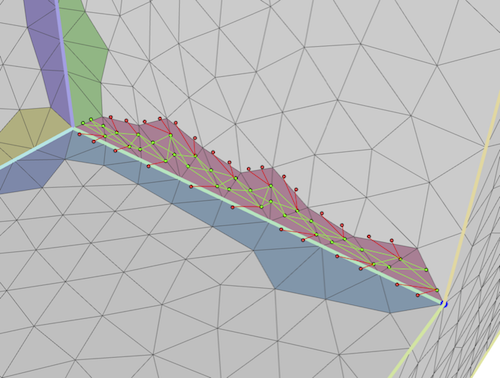} \\
(c)&(d)\\
\end{tabular}
\caption{Tracing loops along curve features requires  modifying the graph $\G$ to avoid the endpoint singularities: we consider the triangles on one side of a curve feature (a), we choose the Steiner points of $\G$ of that side (b) and we weight to zero the red arcs inside the corridor (c), and inhibit the green arcs exiting from the corridor (d). }
\label{fig:sharp_feature_graph}
\end{figure}

\subsection{Extending features to loops}
\label{app:concave-tracing}
In order to force loops to run along line features, we modify the graph $\G$ as follows:
\begin{itemize}
\item Given a line feature $\Lf$, we create two {\em corridors}, each made of a strip of triangles of $\SM$ incident at $\Lf$, one for each side of $\Lf$ (see Figure \ref{fig:sharp_feature_graph}.a);
\item For each face in a corridor, we consider all Steiner nodes of $\G$ that are coherent with the direction of $\Lf$ and that lie on edges crossing the corridor (see Figure \ref{fig:sharp_feature_graph}.b);
\item We reduce the weight of arcs connecting pairs of such nodes in the corridor (green arcs in Figure \ref{fig:sharp_feature_graph}.c);
\item We inhibit all arcs that connect such nodes with nodes at the boundary of the corridor (red arcs in Figure \ref{fig:sharp_feature_graph}.d).
\end{itemize}

For each line feature $\Lf$, we create one seed node per side of $\Lf$ and we trace a set of candidate loops from all such nodes. 
Note that, in the modified graph, each path that enters a corridor is forced to traverse it totally, and paths traversing several corridors (i.e., joining or bridging different line features) are favoured because of their reduced cost (see Figure \ref{fig:multiple_concave_features}.a). 
Note that each feature may be traversed by multiple loops in the set of candidates. 
In the process of generating the loops that extends line features we select loops in a greedy manner, preferring the ones that span the largest length of open features. 

\section{Model cutting}
\label{app:cuts}
\subsection{Similarity metrics for loop pairing}
\label{app:pairing}
We formalize here the similarity metrics for planes and cylinders used to perform loop pairing. Both metrics are defined as penalty metrics in the range $(0,1]$, meaning that lower values denote higher similarity. Given two loops $\ell{i}, \ell{j}$ of type I, II that are centered at $c_i,c_j$ and span two planes with normals $n_i,n_j$, respectively, we define their plane similarity as follows
\[
E_\mathsf{plane} = \frac{1}{2}\left(\exp\left( \frac{- \vert n_i\!\cdot\!n_j \vert^2}{0.2}  \right)
 + \exp\left( \frac{- (1\!-\max(\vert c\!\cdot\!n_i \vert, \vert c\!\cdot\!n_j\vert))^2}{0.2}  \right)\right).
\]
The first term weighs angle similarity between plane normals; the second term weighs the maximum distance between the centroid of $\ell{i}$ and the plane spanned by $\ell{j}$, and vice versa. Vector $c$ is defined as $(c_i-c_j)/\Vert c_i-c_j \Vert$.\\

Given two loops $\ell{i}, \ell{j}$ of type III, that are centered at $c_i,c_j$, have radius $r_i,r_j$ and span two oriented lines $d_i,d_j$, respectively, we define their cylinder similarity as follows
\[
E_\mathsf{cyl} = \frac{1}{2}\left(1-\exp\left( \frac{- (1 + d_i\!\cdot\!d_j)^2}{0.2} \right) + \min \left( \frac{r_i}{r_j},\frac{r_j}{r_i} \right) \right).
\]
The first term measures angle similarity, and promotes pairing between loops laid on oppositely oriented surfaces. The second term matches the radii of the cylinders, and serves to find the best geometric matching in presence of concentric loops.

\subsection{Backup pairing strategy}
\label{app:pairing_backup}
The backup pairing strategy is used when the geometric loop pairing does not provide a globally consistent bi-partition of $\SM$. It computes a bi-partition that assigns to each triangle in the mesh either 0 or 1, and minimizes the total length of the edges having opposite labels at their sides
\[ \min \sum_{e \in \SM} C_e(l_i,l_j) \]
where $e$ are the mesh edges, $l_i,l_j$ are the labels associated to the triangles incident at it, and the per edge cost energy $C_e$ is defined as
\[
C_e(l_i,l_j) = \left\lbrace \begin{array}{cl}
0 & \text{if } l_i = l_j\\
\vert e \vert &  \text{if } l_i \neq l_j\\
\end{array}\right.
\]
We compute the solution using the min-cut implementation provided by~\cite{boykov2001fast,kolmogorov2004energy,boykov2004experimental}. To guarantee that the initial cut is used to bi-partition, we set boundary conditions that impose that the triangles at its two sides must receive opposite labels.

\subsection{Topological cleaning}
\label{app:topoclean}
\begin{figure}[h]
\centering
\includegraphics[width=\linewidth]{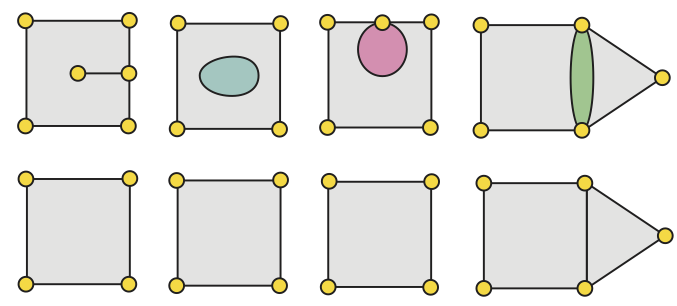}
\caption{Extracting the meta-mesh at early stages of the algorithm may result in cells having a number of topological artifacts, such as: dangling edges (left); islands (middle-left); non manifold vertices (middle-right); and faces having less than three vertices (right). We address such cases during mesh extraction, ignoring topological inconsistencies and collapsing non valid faces (bottom).}
\label{fig:cleaning}
\end{figure}

Since the final meta-mesh reveals itself cut after cut (Figure~\ref{fig:cut_sequence}), extracting it when none or just a few cuts have been performed may result in a number of topological inconsistencies that affect its cells, such as: dangling edges, islands, non-manifold vertices and faces with less than two vertices (Figure~\ref{fig:cleaning}). We make the extraction of the meta-mesh robust against all these defects by using a set of topological operators that collapse all the topologically illegal entities, as depicted in the bottom line of the figure.

\end{document}